\documentclass[useAMS,usenatbib]{aa}  

\usepackage{txfonts}
\usepackage{graphicx}
\usepackage{amsmath}
\usepackage{tabularx}
\usepackage{inputenc}
\usepackage{supertabular}
\usepackage{ltxtable}
\usepackage{subfig}
\usepackage{float}

\begin{document}

   \title{A super-Earth orbiting the nearby M dwarf GJ 536 \thanks{The data used in this paper (Table~\ref{tab_a1}) is available in electronic form at the CDS via anonymous ftp to cdsarc.u-strasbg.fr (130.79.128.5) or via http://cdsweb.u-strasbg.fr/cgi-bin/qcat?J/A+A/}}

   \subtitle{}

      \author{A. Su\'{a}rez Mascare\~{n}o\inst{1,2} \and
         J.~I. Gonz\'{a}lez Hern\'{a}ndez   \inst{1,2} \and
         R. Rebolo  \inst{1,2,3}   \and
         N. Astudillo-Defru \inst{4} \and
         X. Bonfils \inst{5,6} \and
         F. Bouchy \inst{4} \and
         X.Delfosse \inst{5,6} \and 
         T. Forveille \inst{5,6} \and 
         C. Lovis \inst{4} \and 
         M. Mayor \inst{4} \and 
         F. Murgas \inst{5,6} \and 
         F. Pepe \inst{4} \and
         N. C. Santos \inst{7,8} \and
         S. Udry \inst{4} \and
         A. W\"{u}nsche \inst{7,8} \and
         S. Velasco\inst{1,2}
}         
    \institute{Instituto de Astrof\'{i}sica de Canarias, E-38205 La Laguna, Tenerife, Spain\\
              \email{asm@iac.es}          \and
             Universidad de La Laguna, Dpto. Astrof\'{i}sica, E-38206 La Laguna, Tenerife, Spain \and
             Consejo Superior de Investigaciones Cient{\'\i}ficas, Spain \and
             Observatoire Astronomique de l'Université de Genève, Versoix, Switzerland \and 
             Univ. Grenoble Alpes, IPAG, Grenoble, France\and  
             CNRS, IPAG, Grenoble, France\and 
             Instituto de Astrofísica e Ciências do Espa\c{c}o, Universidade do Porto,CAUP, Rua das Estrelas, 4150-762 Porto, Portugal\and       
             Departamento de F\'{i}sica e Astronomia, Faculdade de Ci\^{e}ncias,Universidade do Porto,Rua Campo Alegre, 4169-007  Porto, Portugal             \\
             }

   \date{Written July-2016}

 
  \abstract
   {We report the discovery of a super-Earth orbiting the star GJ 536 based on the analysis of the radial-velocity time series from the HARPS and HARPS-N spectrographs. GJ 536 b is a planet with a minimum mass M sin $i$ of 5.36 $\pm$ 0.69 M$_{\oplus}$; it has an orbital period of 8.7076 $\pm$ 0.0025 days at a distance of 0.066610(13) AU, and an orbit that is consistent with circular. The host star is the moderately quiet M1 V star GJ 536, located at 10 pc from the Sun. We find the presence of a second signal at 43 days that we relate to stellar rotation after analysing the time series of Ca II H\&K and H$_{\alpha}$ spectroscopic indicators and photometric data from the ASAS archive. We find no evidence linking the short period signal to any activity proxy. We also tentatively derived a stellar magnetic cycle of less than 3 years.
}
   
   {}

   \keywords{
              Planetary Systems --- Techniques: radial velocity --- Stars: activity --- Stars: chromospheres --- Stars: rotation --- Stars: magnetic cycle --- starspots --- Stars: individual (GJ 536)
 }

   \maketitle
%

\section{Introduction}

Several surveys have attempted to take advantage of the low masses of M dwarfs  and therefore of the stronger radial-velocity signals induced  for the same planetary mass -- and their closer habitable zones to detect rocky habitable planets \citep{Bonfils2013, Howard2014, Irwin2015, Berta-Thompson2015, AngladaEscude2016}. While surveying M dwarfs has advantages, it also has its own drawbacks. Stellar activity has been one of the main difficulties when trying to detect planets trough Doppler spectroscopy. Not only does it introduce noise, but also coherent signals that can mimic those of planetary origin \citep{Queloz2001, Bonfils2007, Robertson2013}. M dwarfs tend to induce signals with amplitudes comparable to those of rocky planets \citep{Howard2014, Robertson2013}. While these kinds of stars allow for the detection of smaller planets, they also demand a more detailed analysis of the radial-velocity signals induced by activity. In addition, these  low-mass stars offer valuable complementary information on the formation mechanisms of planetary systems. For instance, giants planets are known to be rare around M dwarfs, while    super-Earths appear to be more frequent \citep{Bonfils2013, Dressing2013, Dressing2015}.
 
In spite of the numerous exoplanets detected by \emph{Kepler} \citep{Howard2012} and by radial-velocity surveys \citep{Howard2009, Mayor2011} the number of known small rocky planets is still comparably low. There are around 1500 confirmed exoplanets and more than 3000 Kepler candidates, but only about a hundred of the confirmed planets have been reported on  M dwarfs and only a fraction of them are rocky planets. The first discovery of a planet around an M dwarf dates back to 1998 \citep{Delfosse1998, Marcy1998}. Since then several planetary systems have been reported containing Neptune-mass planets and super-Earths \citep{Udry2007, Delfosse2013, Howard2014,AstudilloDefru2015} and  even some Earth-mass planets \citep{Mayor2009,BertaThompson2015, Wright2015, Affer2016,AngladaEscude2016}. However, the frequency of very low-mass planets around M dwarfs is not well established. In particular, as noted by \citet{Bonfils2013}, the frequency of rocky planets at periods shorter than 10 days is $0.36^{+0.24}_{-0.10}$; it is $0.41^{+0.54}_{-0.13}$ for the habitable zone of the stars. On the other hand \citet{Gaidos2013} estimated that the frequency of habitable rocky planets is $0.46^{+0.20}_{-0.15}$ on a wider spectral sample of Kepler dwarfs, and \citet{Kopparapu2013} gave a frequency of $0.48^{+0.12}_{-0.24}$ for habitable planets around M dwarfs. The three measurements are compatible, but uncertainties are still big making it important to continue the search for planets around this star type  in order  to refine the statistics. 

We present the discovery of a super-Earth orbiting the nearby star GJ 536, which is a high proper motion early M dwarf at a distance of 10 pc from the Sun \citep{vanLeeuwen2007, Maldonado2015}. Because of its high proper motion and its closeness, this star shows a secular acceleration of 0.24 m s$^{-1}$ yr$^{-1}$ \citep{Montet2014}. Table~\ref{parameters} shows the stellar parameters. Its moderately low activity combined with its long rotation period of more than 40 days \citep{Masca2015} makes it a very interesting candidate to search for rocky planets. 

\begin {table}
\begin{center}
\caption { Stellar parameters of GJ 536 \label{tab:parameters}}
    \begin{tabular}{ l  l  l l l l l l l l l l } \hline
Parameter  & GJ 536 & Ref.\\ \hline
RA (J2000) & 14:01:03.19 & 1\\
DEC (J2000) & -02:39:17.52 & 1\\
$\delta$ RA($mas$ yr$^{-1}$) & -823.47 & 1\\
$\delta$ DEC ($mas$ yr$^{-1}$) & 598.19 & 1\\
Distance [pc] & 10.03 & 1 \\
$m_{B}$ & 11.177 & 2\\
$m_{V}$                 & 9.707 & 2 \\  
$m_{V}$ ASAS                    & 9.708 & 0 \\
Spectral Type & M1 & 3\\
T$_{\rm eff}$ [K] & 3685 $\pm$ 68 & 3\\
$[Fe/H]$ & -0.08 $\pm$ 0.09 & 3\\
$M_\star$~$[M_{\odot}]$ & 0.52 $\pm$ 0.05 & 3\\
$R_\star$~$[R_{\odot}]$ & 0.50 $\pm$ 0.05 & 3\\
log $g$ (cgs) &4.75 $\pm$ 0.04 & 3\\
log($L_{\star}/L_{\odot}$) & -1.377 & 3 \\
$\log_{10}(R'_\textrm{HK})$ &  -5.12 $\pm$ 0.05 & 0\\
$P_{\rm rot}$ & 45.39 $\pm$ 1.33 & 0\\
$v \sin i$ (km s$^{-1}$) & $\textless$ 1.2$^{*}$ &  0\\
Secular acc. (m s$^{-1}$ yr$^{-1}$) & 0.24 & 4 \\
 \hline
\label{parameters}
\end{tabular}  
\end{center}
\textbf{References:} 0 - This work, 1 - \citet{vanLeeuwen2007}, 2 - \citet{Koen2010}, 3 -\citet{Maldonado2015}, 4 - Calculated following \citet{Montet2014}.\\
$^{*}$ Estimated using the Radius estimated by \citet{Maldonado2015} and our period determination. \\
\end {table}

The star GJ 536 is part of the \citet{Bonfils2013} sample and has been extensively monitored since mid-2004. We have used 146 HARPS spectra taken over 11.7 yr along with 12 HARPS-N spectra taken during April and May 2016. HARPS \citep{Mayor2003} and HARPS-N \citep{Cosentino2012} are two fibre-fed high-resolution echelle spectrographs installed at the 3.6 m ESO telescope in La Silla Observatory (Chile) and at the Telescopio Nazionale Galileo in the Roque de los Muchachos Observatory (Spain), respectively. Both instruments have a resolving power greater than $R\sim 115\,000$ over a spectral range from $\sim$380 to $\sim$690 nm and have been designed to attain very high long-term radial-velocity accuracy. Both are contained in vacuum vessels to avoid spectral drifts due to temperature and air pressure variations, thus  ensuring their stability. HARPS and HARPS-N are equipped with their own pipeline providing extracted and wavelength-calibrated spectra, as well as RV measurements and other data products such as cross-correlation functions (CCFs) and their bisector profiles. 

Most of the observations were carried out using the Fabry P\'erot  interferometer (FP) as simultaneous calibration. The FP offers the possibility of monitoring the instrumental drift with a precision of 10 cms$^{-1}$ without the risk of contamination of the stellar spectra by the ThAr saturated lines \citep{Wildi2010}. While this is not usually a problem in G and K stars, the small amount of light collected in the blue part of the spectra of M dwarfs might compromise the quality of the measurement of the Ca II H\&K flux.  The FP allows a precision of $\sim$ 1 ms$^{-1}$ in the determination of the radial velocities of the spectra with highest signal-to-noise ratios while assuring the quality of the spectroscopic indicators even in those spectra with low signal-to-noise ratios. Measurements taken before the availability of the FP were taken without simultaneous reference.

We also use the photometric data on GJ 536 provided by the All Sky Automated Survey (ASAS) public database. ASAS \citep{Pojmanski1997} is an all-sky survey in the $V$ and $I$ bands running since 1998 at Las Campanas Observatory, Chile.  The best photometric results are achieved for stars with V $\sim$8-14, but this range can be extended  by  implementing  some quality control on the data. ASAS has produced  light curves for around $10^{7}$ stars at $\delta < 28^{\circ}$. The ASAS catalogue supplies ready-to-use light curves with flags indicating the quality of the data. For this analysis we relied  only on  good quality data (grade A and B in the internal flags). Even after this quality control,  there are still some high dispersion measurements  which cannot  be explained by a regular stellar behaviour. We reject those measurements by de-trending the series and eliminating points deviating more than three times the standard deviation from the median seasonal value. We are left with 359 photometric observations taken over 8.6 yr with a typical uncertainty of 9.6 mmag per exposure.

\section{Determining  stellar activity indicators and radial velocities}

\subsection{Activity indicators}
For the activity analysis we use the extracted order-by-order wavelength-calibrated spectra produced by the HARPS and HARPS-N pipelines. For a given star, the change in atmospheric transparency from day to day causes variations in the  flux distribution of the recorded spectra that are particularly relevant in the blue where we intend to measure Ca II lines. In order to minimize  the effects related to these atmospheric changes  we create a spectral template for each star by de-blazing and co-adding every available spectrum; we  use the co-added spectrum to correct the order-by-order flux of each individual spectrum. We also   correct each spectrum for the Earth's barycentric radial velocity and the radial velocity of the star using the measurements given by the standard pipeline, and re-binned the spectra into a wavelength-constant step. Using this  HARPS dataset, we expect to have high-quality spectroscopic indicators  to monitor tiny stellar activity variations with  high accuracy. 

\subsection*{S$_{MW}$ index}
\begin{figure}
\centering
        \includegraphics[width=9.0cm]{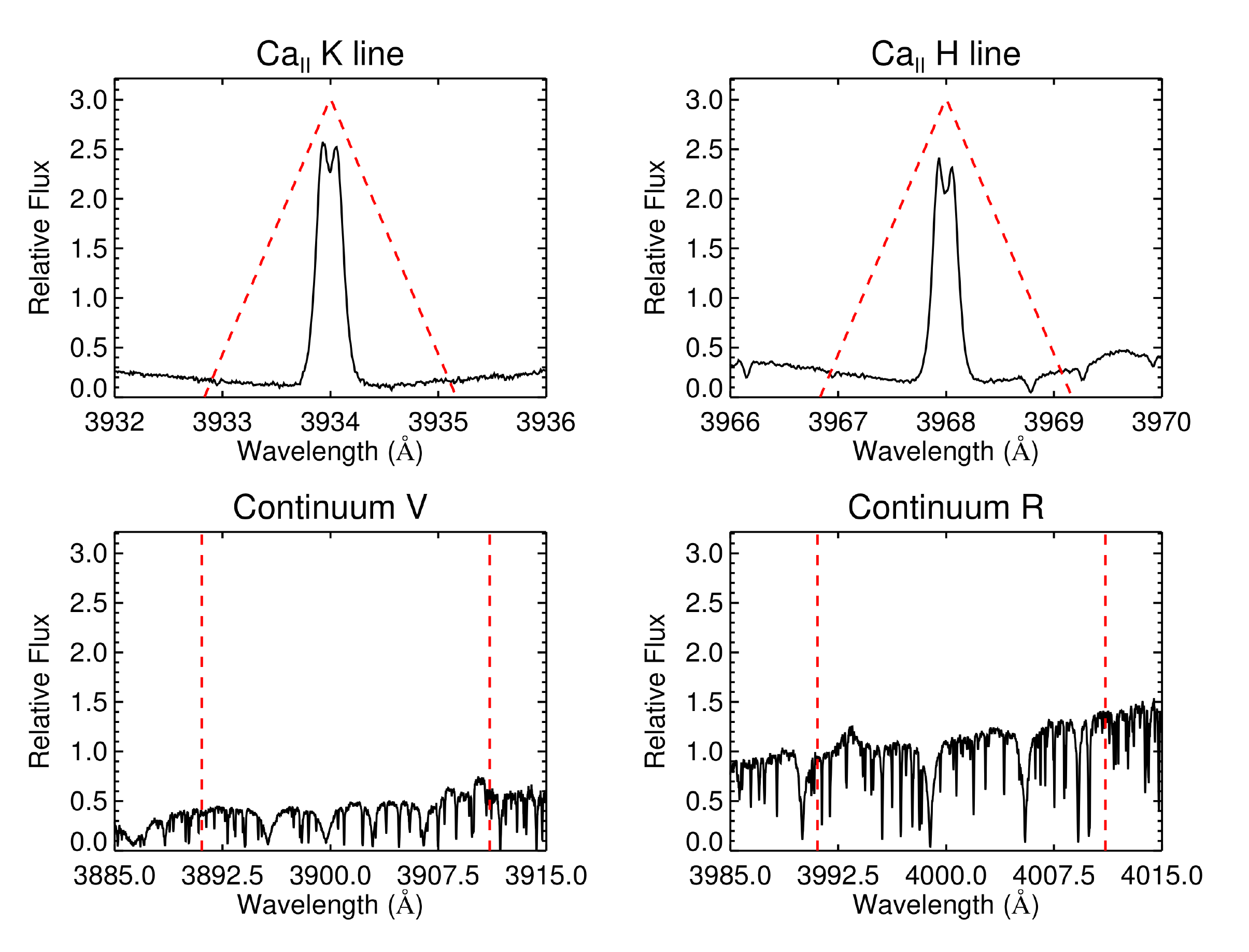}
        \caption{Ca II H\&K filter of the spectrum of the star GJ536 with the same shape as the Mount Wilson Ca II H\&K passband.}
        \label{Sindex}
\end{figure}

We calculate the Mount Wilson $S$ index and the $\log_{10}(R'_{HK})$ by using the original \citet{Noyes1984} procedure, following \citet{Lovis2011} and \citet{Masca2015}. We define two triangular passbands with  full width half maximum (FWHM) of 1.09~{\AA} centred at 3968.470~{\AA} and 3933.664~{\AA} for the Ca II H\&K line cores, and for the continuum we use two 20~{\AA} wide bands centred at 3901.070~{\AA} (V) and 4001.070~{\AA}(R), as shown in Fig.~\ref{Sindex}.

Then the S-index is defined as 

\begin{equation}
   S=\alpha {{\tilde{N}_{H}+\tilde{N}_{K}}\over{\tilde{N}_{R}+\tilde{N}_{V}}} + \beta,
\end{equation}
\noindent where $\tilde{N}_{H},\tilde{N}_{K},\tilde{N}_{R}$, and $\tilde{N}_{V}$ are the mean fluxes in each passband,  while $\alpha$ and $\beta$ are calibration constants fixed as $\alpha = 1.111$ and $\beta = 0.0153$ . The S index serves as a measurement of the Ca II H\&K core flux normalized to the neighbour continuum. As a normalized index to compare it to other stars, we compute the $\log_{10}(R'_\textrm{HK})$ following \citet{Masca2015}.

\subsection*{H$_{\alpha}$ index}

\begin{figure}
        \includegraphics[width=9.0cm]{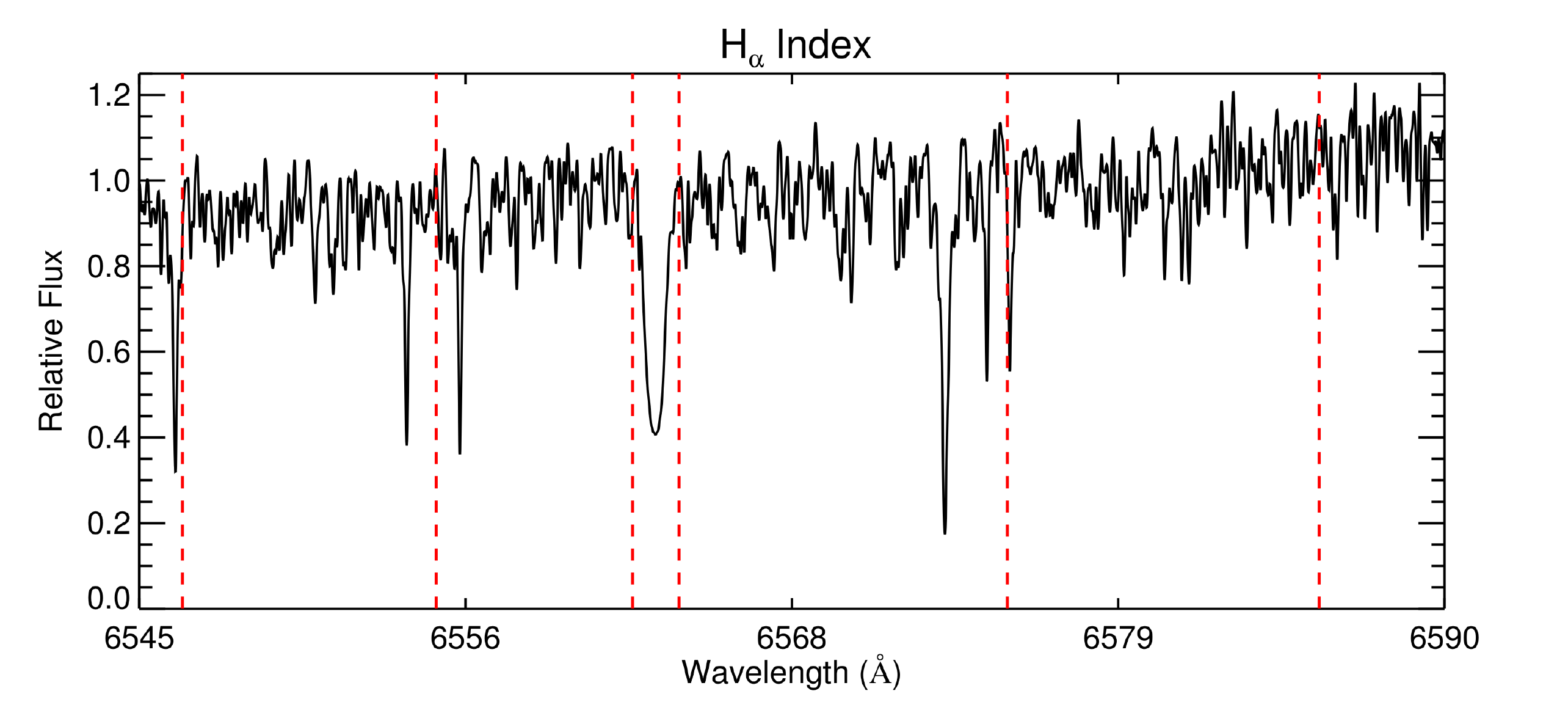}
        \caption{Spectrum of the M-type star GJ536 showing the H$\alpha$ filter passband and continuum bands.}
        \label{halpha}
\end{figure}

We also use the H$\alpha$ index, with a simpler passband following \cite{GomesdaSilva2011}. It consists of a rectangular bandpass with a width of 1.6~{\AA} and centred at 6562.808~{\AA} (core), and two continuum bands of 10.75~{\AA} and  8.75~{\AA} in width centred at 6550.87~{\AA} (L) and 6580.31~{\AA} (R), respectively, as seen in Figure~\ref{halpha}. 

Thus, the H$\alpha$ index is defined as

\begin{equation}
   H\alpha_{\rm Index}={{H\alpha_{\rm core} }\over{H\alpha_{L} +H\alpha_{R}}}.
\end{equation}

\subsection{Radial velocities}

The radial-velocity measurements in the HARPS standard pipeline is determined by a Gaussian fit of the  CCF of the spectrum with a binary stellar template \citep{Baranne1996, PepeMayor2000}. In the case of M dwarfs, owing to the huge number of line blends, the CCF is not Gaussian and results in a less precise Gaussian fit which might cause distortions in the radial-velocity measurements and FWHM. To deal with this issue we tried two different approaches. 

\begin{figure}
        \includegraphics[width=9.0cm]{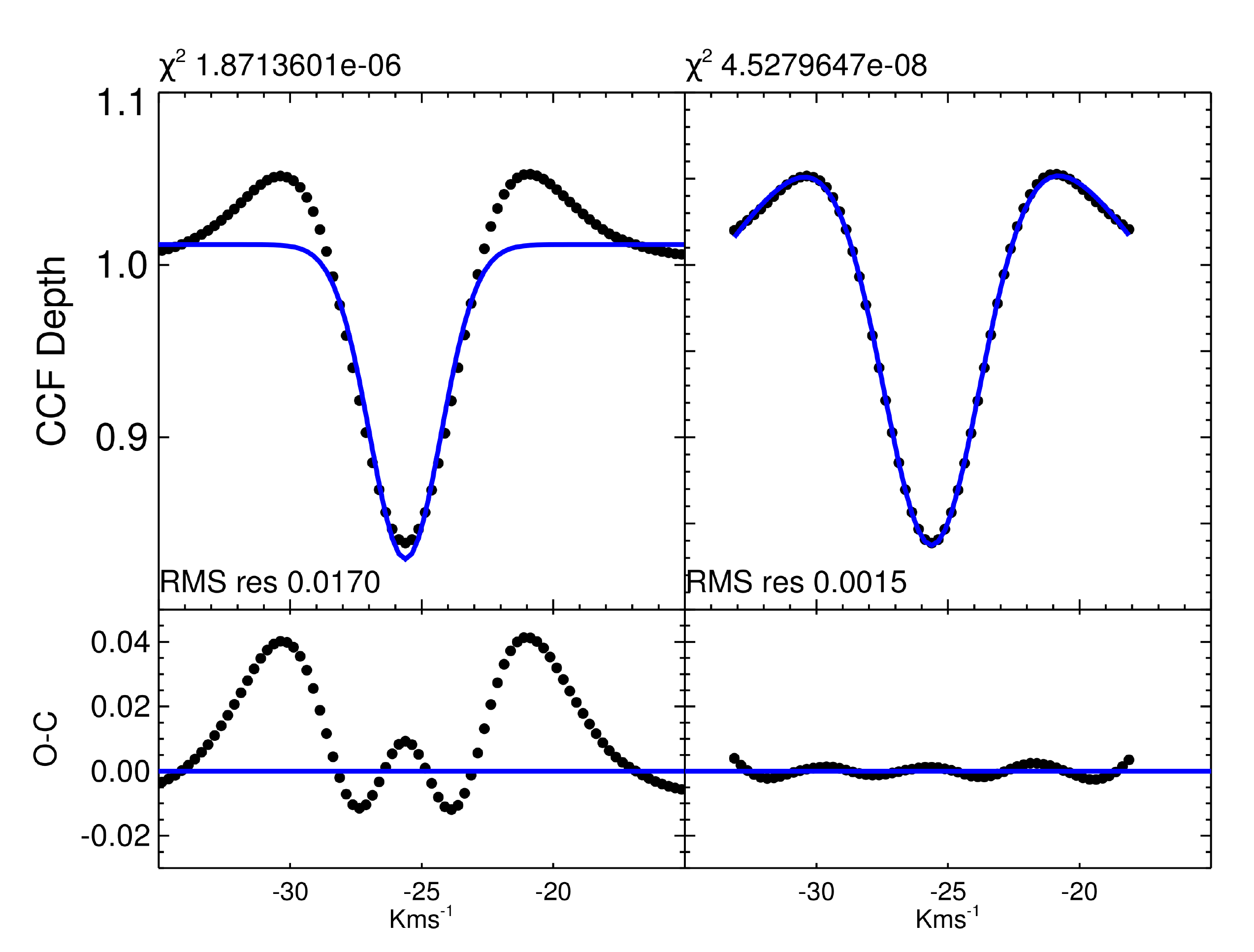}
        \caption{Cross-correlation function for GJ 536. Upper panels show the CCF with the Gaussian fit (left) and our Gaussian plus polynomial fit (right). Lower panels show the residuals after the fit for the Gaussian fit (left) and for our fit. Blue lines show the fit (upper panels) and the zero line (lower panels).}
        \label{ccf_plot}
\end{figure}

The first approach consisted in using a slightly more complex model for the CCF fitting, a Gaussian function plus a second-order polynomial (Fig.~\ref{ccf_plot}) using only the central region of the CCF function. We use a 15 Km s$^{-1}$ window centred at the minimum of the CCF. This configuration provides the best stability of the measurements. Along  with the measurements of the radial velocity we obtain the FWHM of the cross-correlation function, which we also use to track variations in the  activity level of the star. The second approach to the problem was to recompute the radial velocities using a template matching algorithm with a high signal-to-noise stellar spectral template \citep{AstudilloDefru2015}. Every spectrum is corrected from both barycentric and stellar radial velocity to align it to the frame of the solar system barycentre. The radial velocities are computed by minimizing the $\chi^2$ of the residuals between the observed spectra and shifted versions of the stellar template, with all the elements contaminated by telluric lines masked. All radial-velocity measurements are corrected from the secular acceleration of the star.

For the bisector span measurement we rely on the pipeline results as it does not depend on the fit, but on the CCF itself. The bisector has been  a standard activity diagnostic tool for solar-type stars for
more than 10 years. Unfortunately, its behaviour in slow rotating stars is not as informative as it is for fast rotators \citep{Saar1997, Bonfils2007}. We report the measurements of the bisector span (BIS) for each radial-velocity measurement, but we do not find any meaningful information in its analysis.

\subsection{Quality control of the data}

As the sampling rate of our data is not  well suited for  modelling fast events, such as flares, and their effect on the radial velocity is not well understood, we identify and reject points likely affected by flares by searching for  abnormal behaviour of the activity indicators \citep{Reiners2009}. The process rejected six spectra that correspond to flare events of the star with obvious activity enhancement and line distortion. That leaves us with 140 HARPS spectroscopic observations taken over 10.7 years, with most of the measurements taking place after 2013, with  a typical exposure of 900 s and an average signal-to-noise ratio of 56 at $5500$~{\AA}. We do not apply the quality control procedure to the HARPS-N data as the number of spectra is not high enough.

\section{Stellar activity analysis}

In order to properly understand the behaviour of the star, our first step is to analyse the different modulations present in the photometric and spectroscopic time-series. 

We search for periodic variability compatible with both stellar rotation and long-term magnetic cycles. We compute the power spectrum using a generalised Lomb-Scargle periodogram \citep{Zechmeister2009} and if there is any significant periodicity we fit the detected period using a sinusoidal model, or a double sinusoidal model to account for the asymmetry of some signals \citep{BerdyuginaJarvinen2005} with the MPFIT routine \citep{Markwardt2009}. 

The significance of the periodogram peak is evaluated using  the \citet{Cumming2004} modification of the \citet{HorneBaliunas1986} formula to obtain the spectral density thresholds for the  desired false alarm probability (FAP) levels and the bootstrap randomization \citep{Endl2001} of the data.

\begin{figure}
        \includegraphics[width=9.0cm]{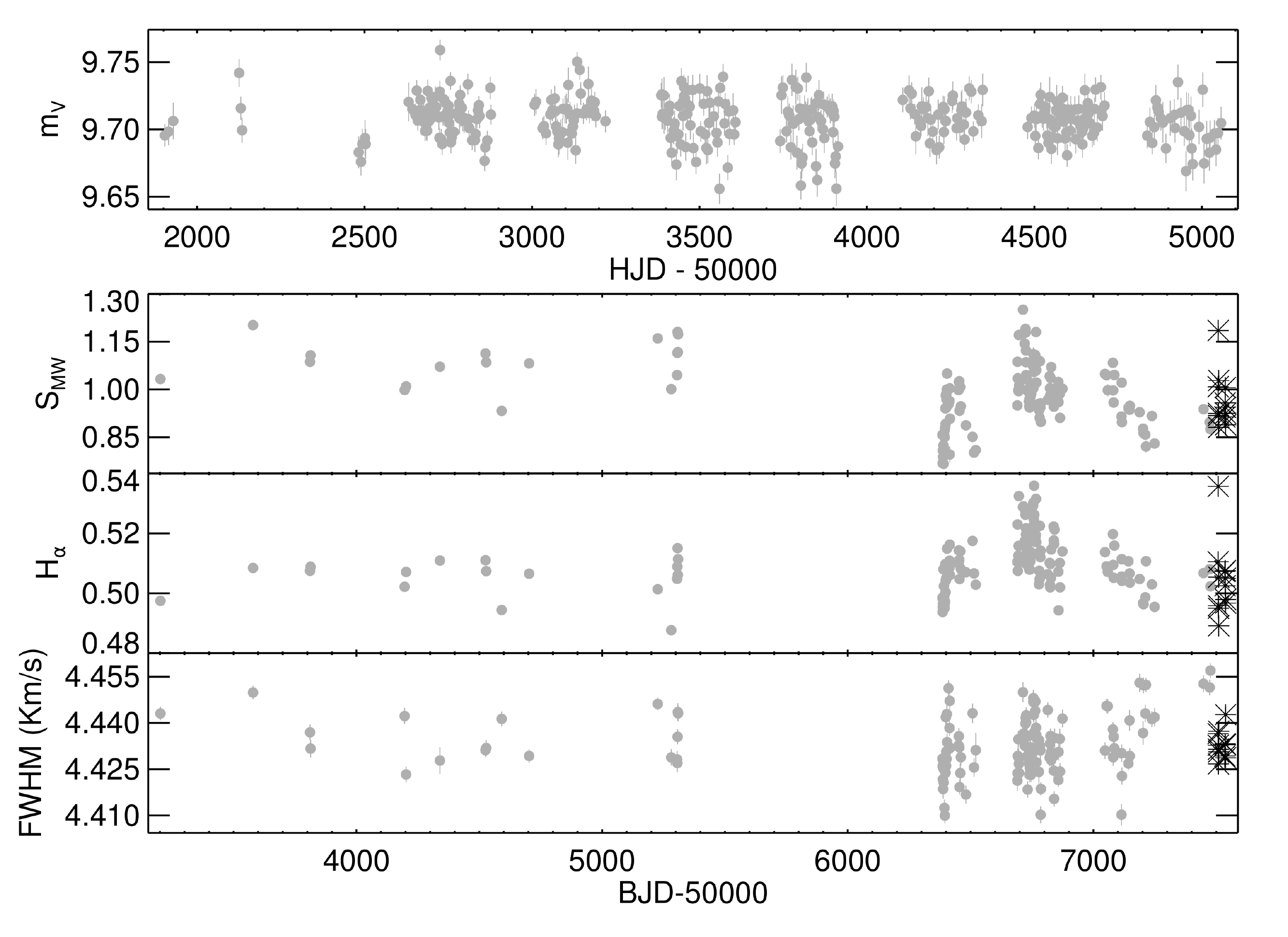}
        \caption{From top  to bottom: Time series of the m$_{V}$, S$_{MW}$ index, H$_{\alpha}$ index, and FWHM time series. Grey dots show HARPS-S data; black asterisks show HARPS-N data.}
        \label{GJ_ac_data}
\end{figure}

Figure~\ref{GJ_ac_data} shows the time series for the photometry (top panel)  and the three activity proxies (bottom panels) used for this analysis. The periodograms of both the photometric and FWHM time series show significant signals at $\sim$40 days, compatible with the typical rotation periods of low-activity M1 stars \citep{Masca2016a, Newton2016}. On the other hand, the periodograms of the S$_{MW}$ and H$_{\alpha}$ indexes show long-term and short-term significant signals. The short period signal is again at $\sim$40 days, while the long-term signal is close to $\sim$1000 days. 

 \begin{figure}
        \includegraphics[width=9.0cm]{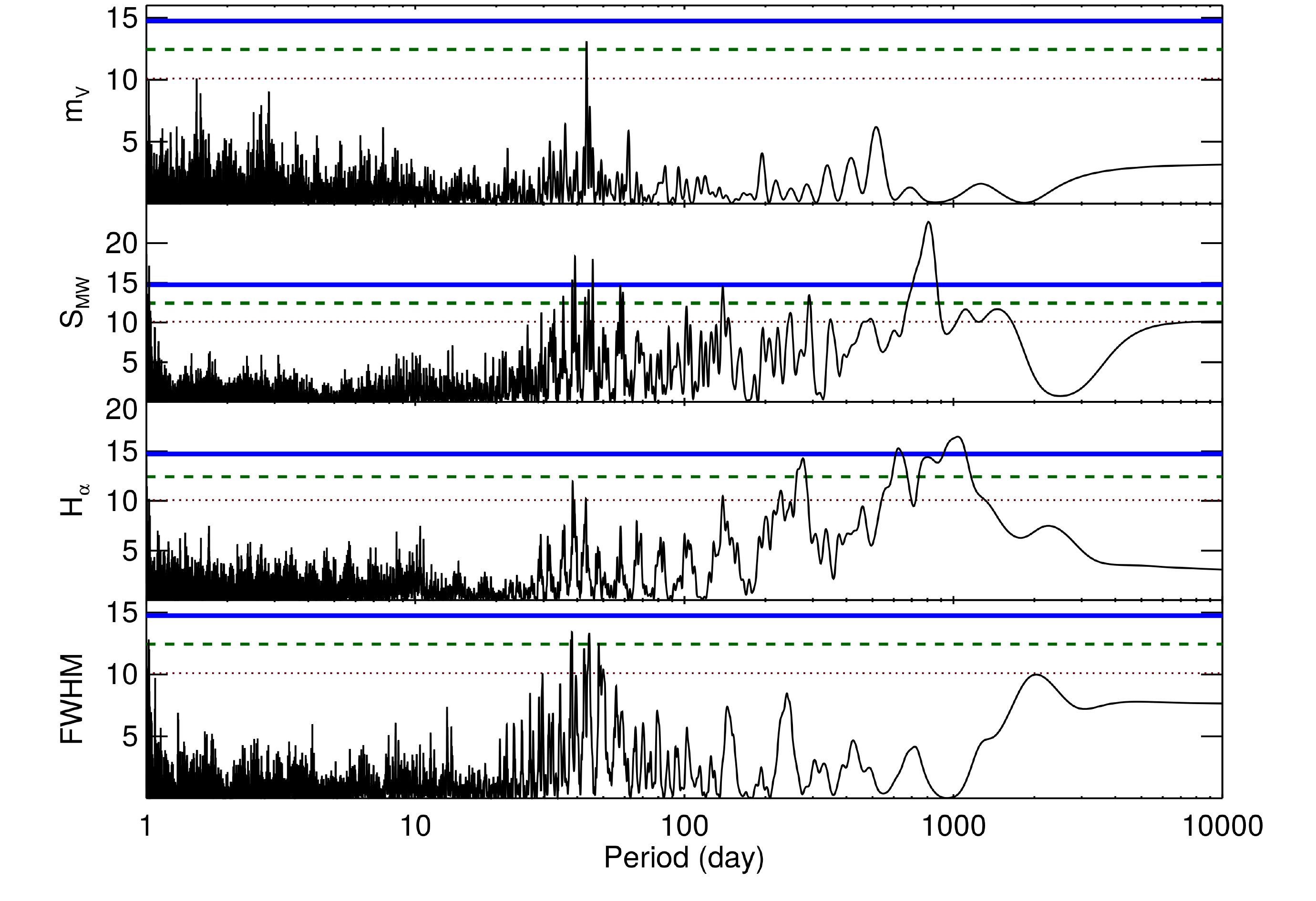}
        \caption{From top to bottom: Periodograms of the m$_{V}$, S$_{MW}$ index, H$_{\alpha}$ index, and FWHM time series. Horizontal lines show the different levels of false alarm probability. Red dotted line for a 10\%  false alarm probability, green dashed line for  1\%, and blue thick line for  0.1\%. Several peaks arise with significances better than  0.1\%.}
        \label{LT_period}
\end{figure}

\subsection{Long-term magnetic cycle}

\begin{figure*} 
        \begin{minipage}{0.5\textwidth}
        \centering
       \includegraphics[width=9.0cm]{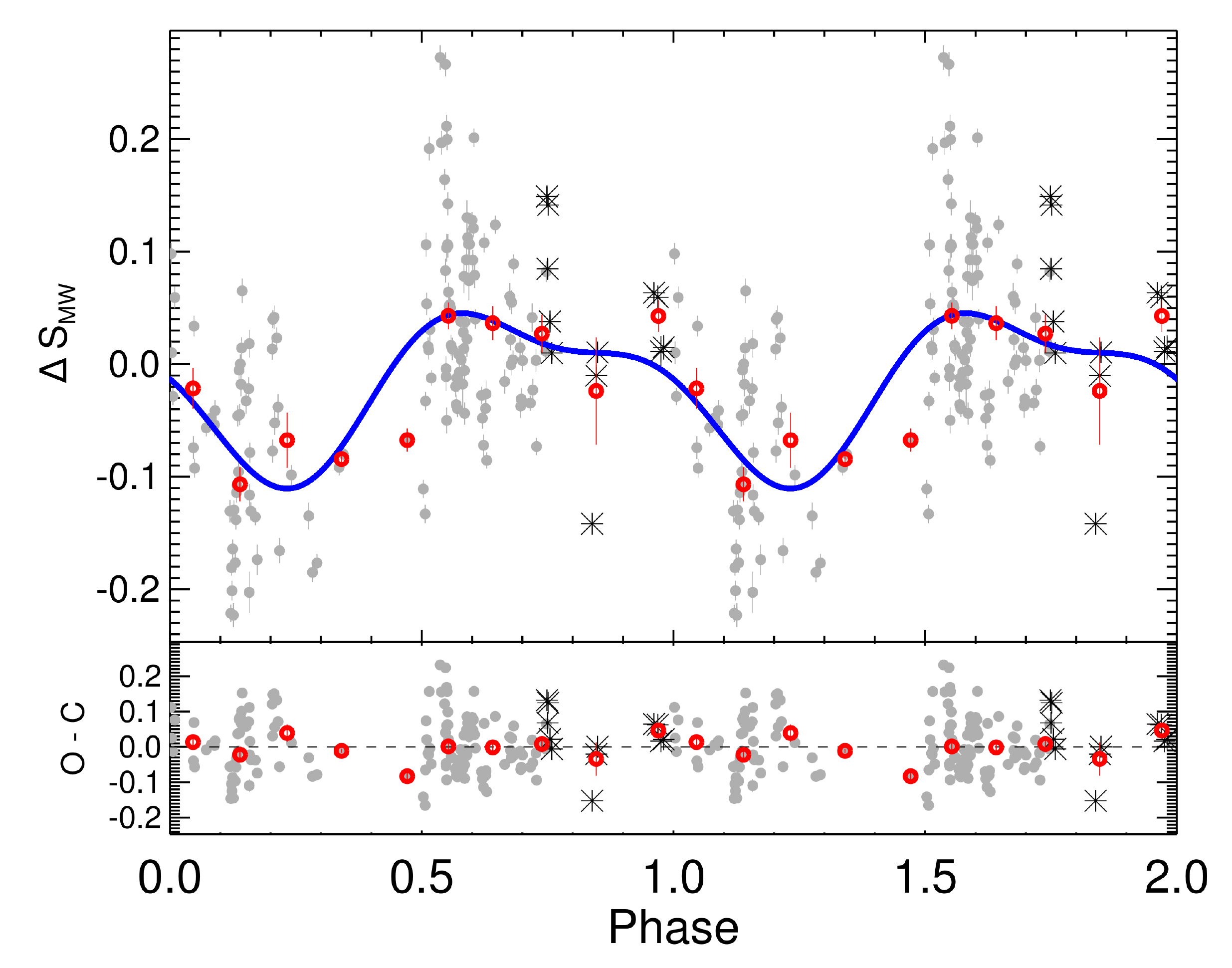}
\end{minipage}%
\begin{minipage}{0.5\textwidth}
        \centering
        \includegraphics[width=9.0cm]{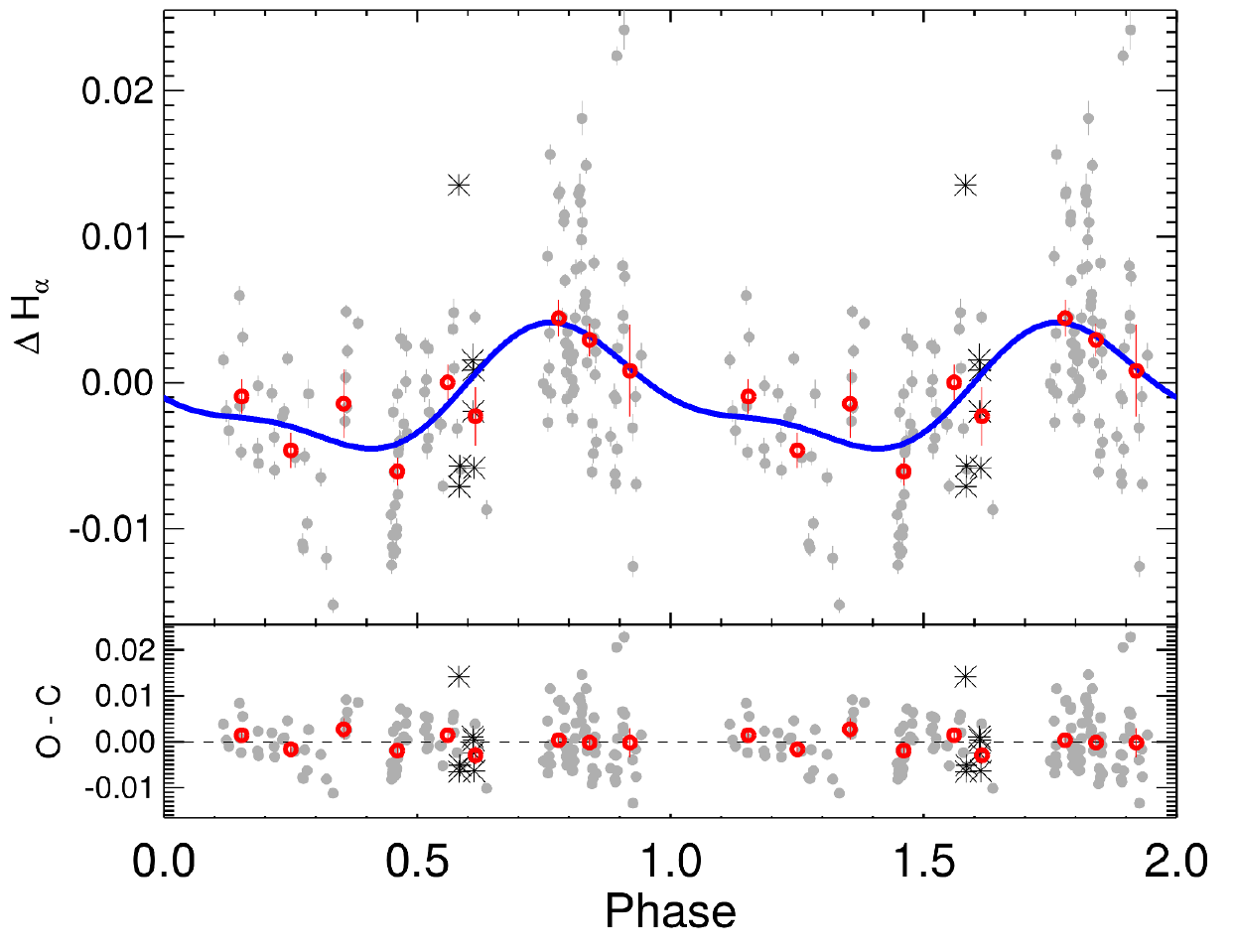}
\end{minipage}%
        \caption{Phase folded fit for the isolated long period activity signal using double-harmonic sine curves.  The left panel shows the S$_{MW}$ index data using the 824 d signal, while the right panel shows the H$_{\alpha}$ signal using the 1075 d signal. Grey dots are the raw measurements after subtracting the mean value. Red dots are the same points binned in phase with a bin size of 0.1.  }
        \label{LT_fit}
\end{figure*}

Analysing the S$_{MW}$ and H$_{\alpha}$ indexes time series we find the presence of a  long-term magnetic cycle of $\sim$3 years. Figure~\ref{LT_period} shows the periodograms of the time series of both indexes. We see a well-defined peak in the  S$_{MW}$ index periodogram at $\sim$806 d and several peaks going from $\sim$600 d to 1100 d in the H$_{\alpha}$ index periodogram implying that the shape of the cycle is still not well defined within our observations. Table~\ref{periodicities} shows the periods of the best fits for both time series using least-squares minimization with the period corresponding to the highest peak of the periodogram as the initial guess. Figure~\ref{LT_fit} shows the phase folded curves  using these periods. The two estimates differ significantly. This might be because of a sub-optimal sampling to detect signals of long periods. The detected periodicities might not be the true periodicities, but apparent periodicities close to the real one caused by the sampling. This also makes us think that the uncertainties in the cycle length are underestimated. The length of the signal is shorter than the typical magnetic cycles measured in solar-type stars, but is within the range of known magnetic cycles in M-type stars \citep{Masca2016a}. In   the S$_{MW}$ and the H$_{\alpha}$ indexes it seems that the cycle shape shows a quick rise followed by a slow decline, as  is the case in the Sun and many other main sequence stars \citep{Waldmeier1961, Baliunas1995}. Unfortunately this cycle is not well covered in phase, making it difficult  to properly characterise it. More observations are needed in order to better constrain its period.

\subsection{Rotation}

The other activity signal expected in our data is the rotational modulation of the star. It shows up at $\sim$43 d with a false alarm probability close to or smaller than  1\% in the four time series (Fig.~\ref{LT_period}) that grow in significance after removing the long-term effects. 

\begin{figure*}

\begin{minipage}{0.5\textwidth}
        \centering
        \includegraphics[width=9.0cm]{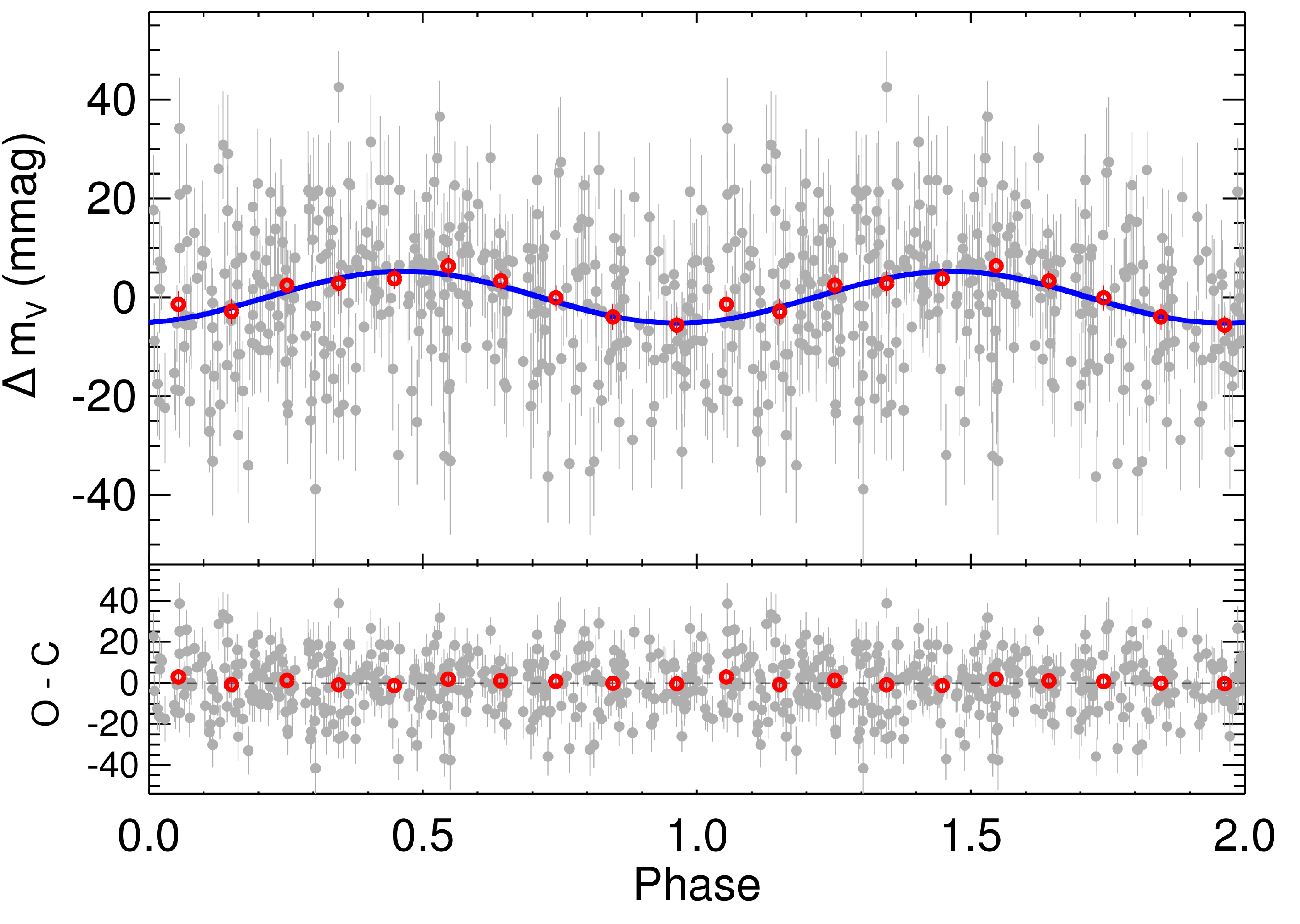}
\end{minipage}%
\begin{minipage}{0.5\textwidth}
        \centering
        \includegraphics[width=9.0cm]{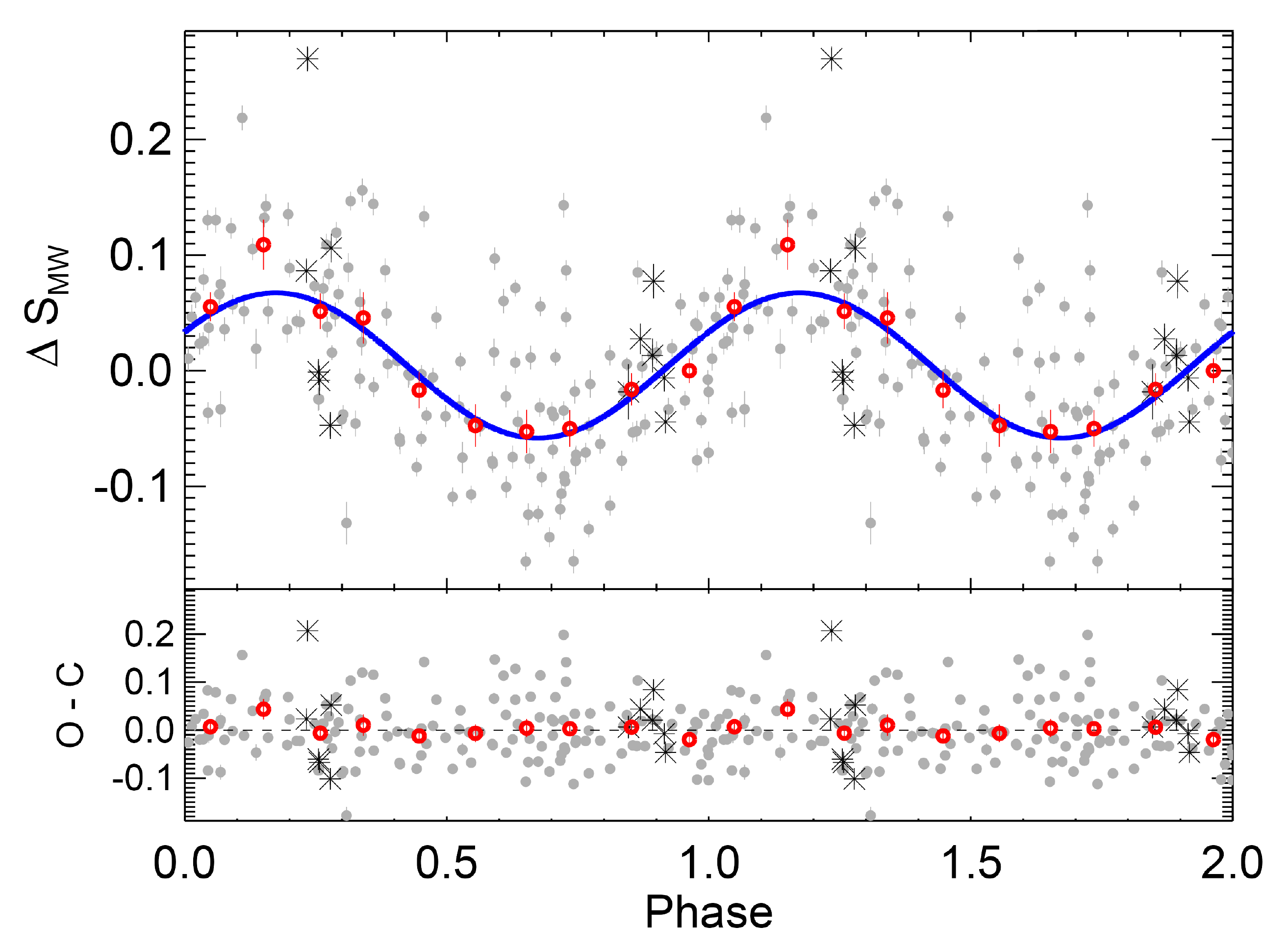}
\end{minipage}%
    
\begin{minipage}{0.5\textwidth}
        \centering
        \includegraphics[width=9.0cm]{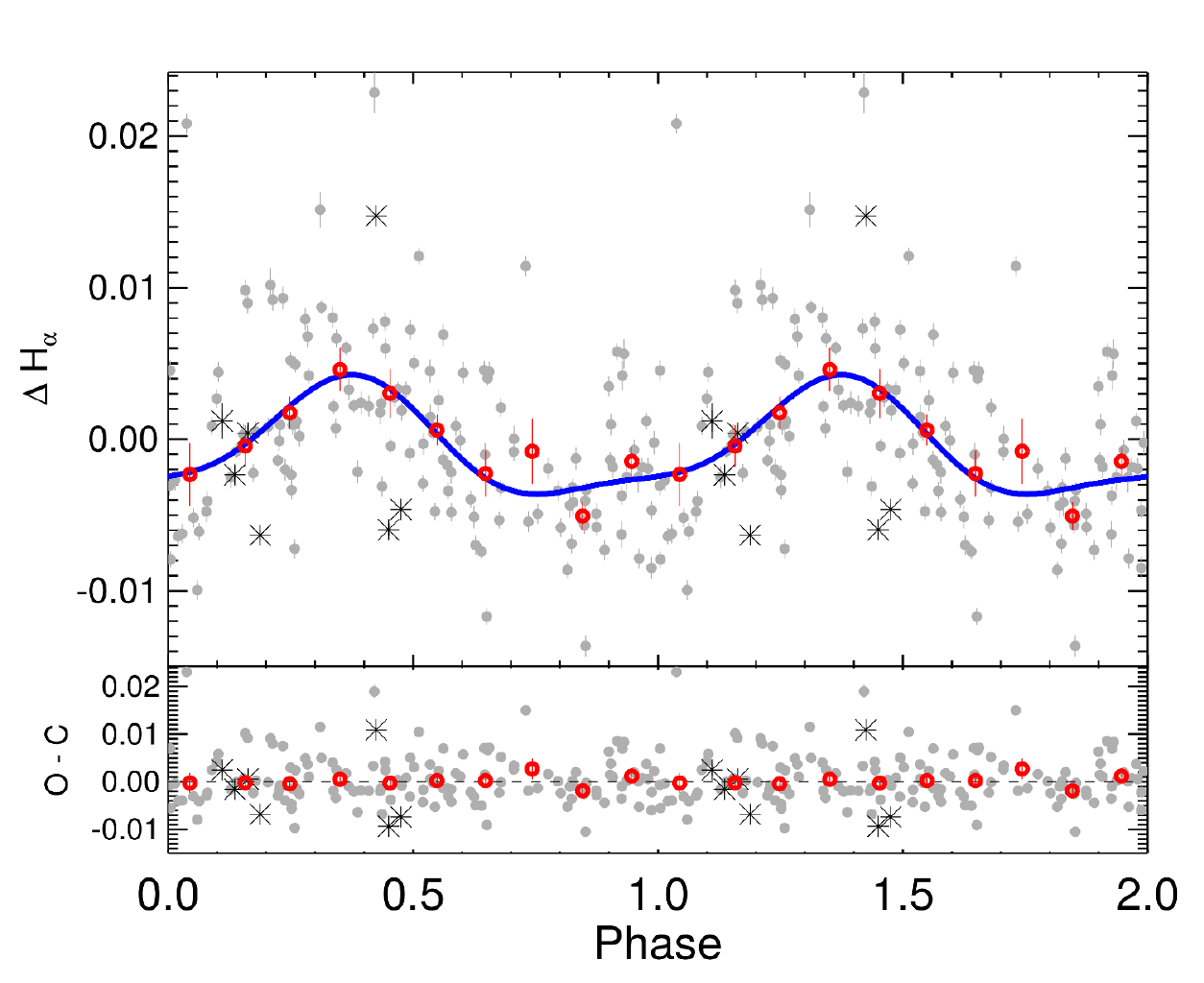}
\end{minipage}%
\begin{minipage}{0.5\textwidth}
        \centering
        \includegraphics[width=9.0cm]{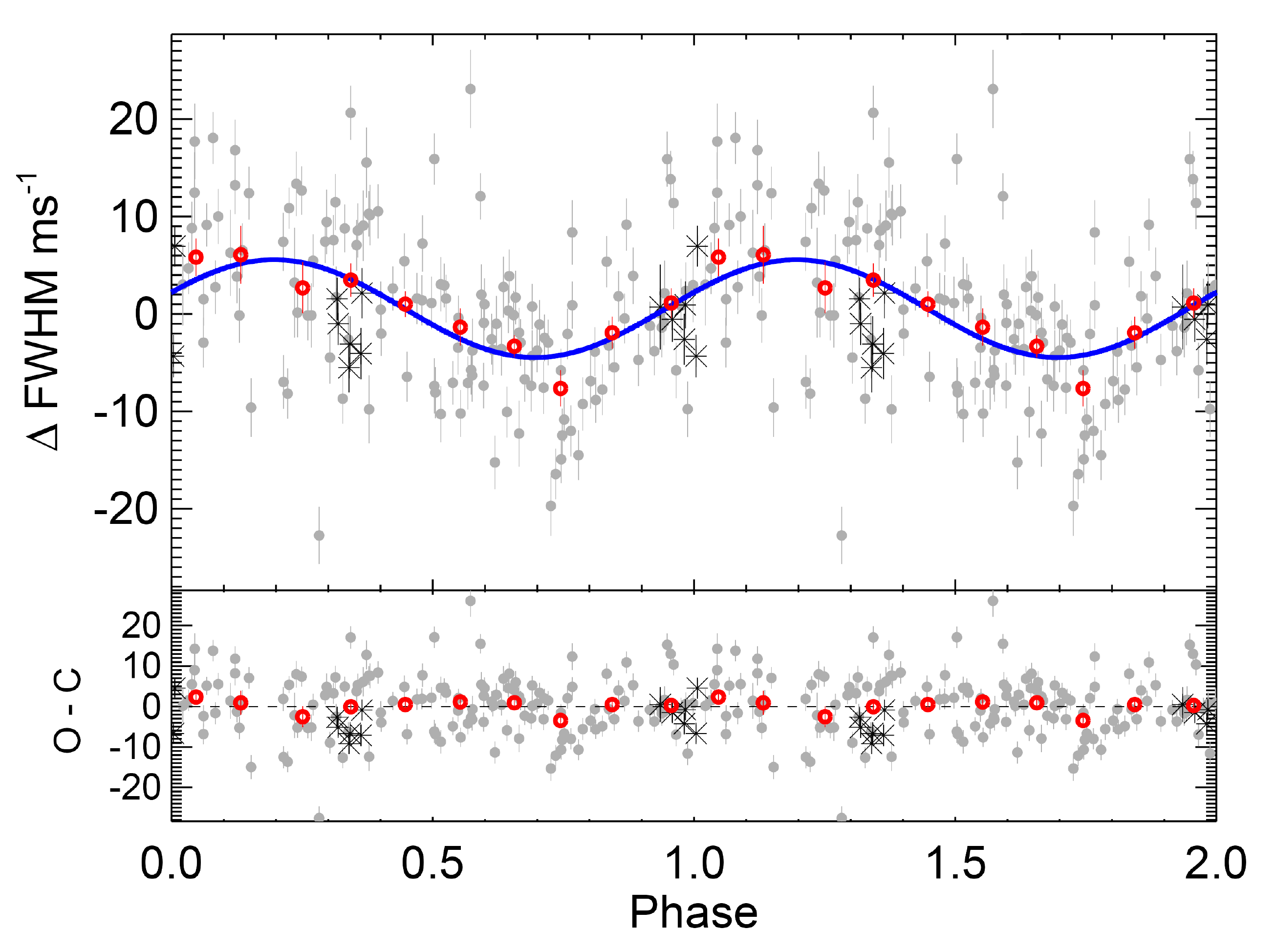}
\end{minipage}%
        \caption{Phase folded curve using the rotational modulation for the ASAS light curve (upper left), S$_{MW}$ index (upper right), H$_{\alpha}$ index using a double-harmonic sine curve (lower left), and FWHM (lower right).  Grey dots are the raw measurements after subtracting the mean value. Red dots are the same points binned in phase with a bin size of 0.1. The error bar of a given bin is estimated using the weighted standard deviation of binned measurements divided by the square root of the number of measurements included in this bin. This estimation of the bin error bars assumes white noise, which is justified by the binning in phase, which regroups points that are uncorrelated in time. }
        \label{rot_phase}
\end{figure*}

In the photometric light curve we measure a  modulation of 43.33 $\pm$ 0.06 d with an amplitude of 5.21 $\pm$ 0.68 mmag. For the S$_{MW}$ index we find a signal of 43.84 $\pm$ 0.01 d with an amplitude of 0.0628 $\pm$ 0.0010 when doing a simultaneous fit with the 824-day signal from Table~\ref{LT_fit}. In the case of the H$_{\alpha}$ index we find a signal 42.58 $\pm$ 0.08 d with an amplitude of 0.0042 $\pm$ 0.0010, also when doing a simultaneous fit with the $\sim$1075 days signal. The time series of the FWHM show a linear increase with time of $\sim$ 2 ms$^{-1}$yr$^{-1}$, which might be related to a slow focus drift of HARPS. After subtracting the linear trend we again find  a periodicity of 44.47 $\pm$ 0.03 d period with an amplitude of 4.56 $\pm$ 0.31 ms$^{-1}$. Figure~\ref{rot_phase} shows the phase folded fits of all the signals. The S$_{MW}$ index and FWHM signals seem to be in phase, while the photometric signal is shifted by a quarter  phase. The uncertainty in the H$_{\alpha}$ long-term fit makes it difficult to give it a unique phase to the rotation signal. Table~\ref{periodicities} shows the parameters for the four signals. 

Our measurement of 45.39 d strengthens the previous estimation of \citet{Masca2015}. Having such a clear detection of the rotational modulation in that many indicators over so many years supports the idea that activity regions in at least some M-type stars are stable over long time spans \citep{Robertson2015}.

\begin {table}
\begin{center}
\caption { Magnetic cycle and rotation periodicities \label{tab:periodicities}}
    \begin{tabular}{ l  l  l l l l l l l l l l } \hline
Series  & Period (d)  & Amplitude &  FAP (\%) \\ \hline
S$_{MW ~ \rm Cyc}$                      & 824.9 $\pm$ 1.7 & 0.0684 $\pm$ 0.0011 & $\textless$ 0.1\\
H$_{\alpha ~ \rm Cyc}$          & 1075.8 $\pm$ 36.1     & 0.0046 $\pm$ 0.0011 & $\textless$ 0.1\\ 
 \\
m$_{V ~ \rm Rot}$                       & 43.33 $\pm$ 0.06      & 5.21 $\pm$ 0.68 mmag & $\textless$ 1\\     
S$_{MW ~ \rm Rot}$                      & 43.84 $\pm$ 0.01      & 0.0628 $\pm$ 0.0010 & $\textless$ 0.1\\
H$_{\alpha ~ \rm Rot}$          & 42.58 $\pm$ 0.08      & 0.0042 $\pm$ 0.0010  & $\textless$ 0.1\\ 
FWHM$_{\rm Rot} $               & 44.47 $\pm$ 0.03      & 4.56 $\pm$ 0.31 ms$^{-1}$  & $\textless$ 1\\ \\
\textbf{$\textless$  Rot. $\textgreater$ }                      & \textbf{43.87 $\pm$ 0.80} \\
\hline

\label{periodicities}
\end{tabular}  
\end{center}
The mean value is the weighted mean of all the individual measurements. The error of the mean is the standard deviation of the individual measurements divided by the square root of the number of measurements.
\end {table}

\section{Radial-velocity analysis}

\begin{figure}
        \includegraphics[width=9.0cm]{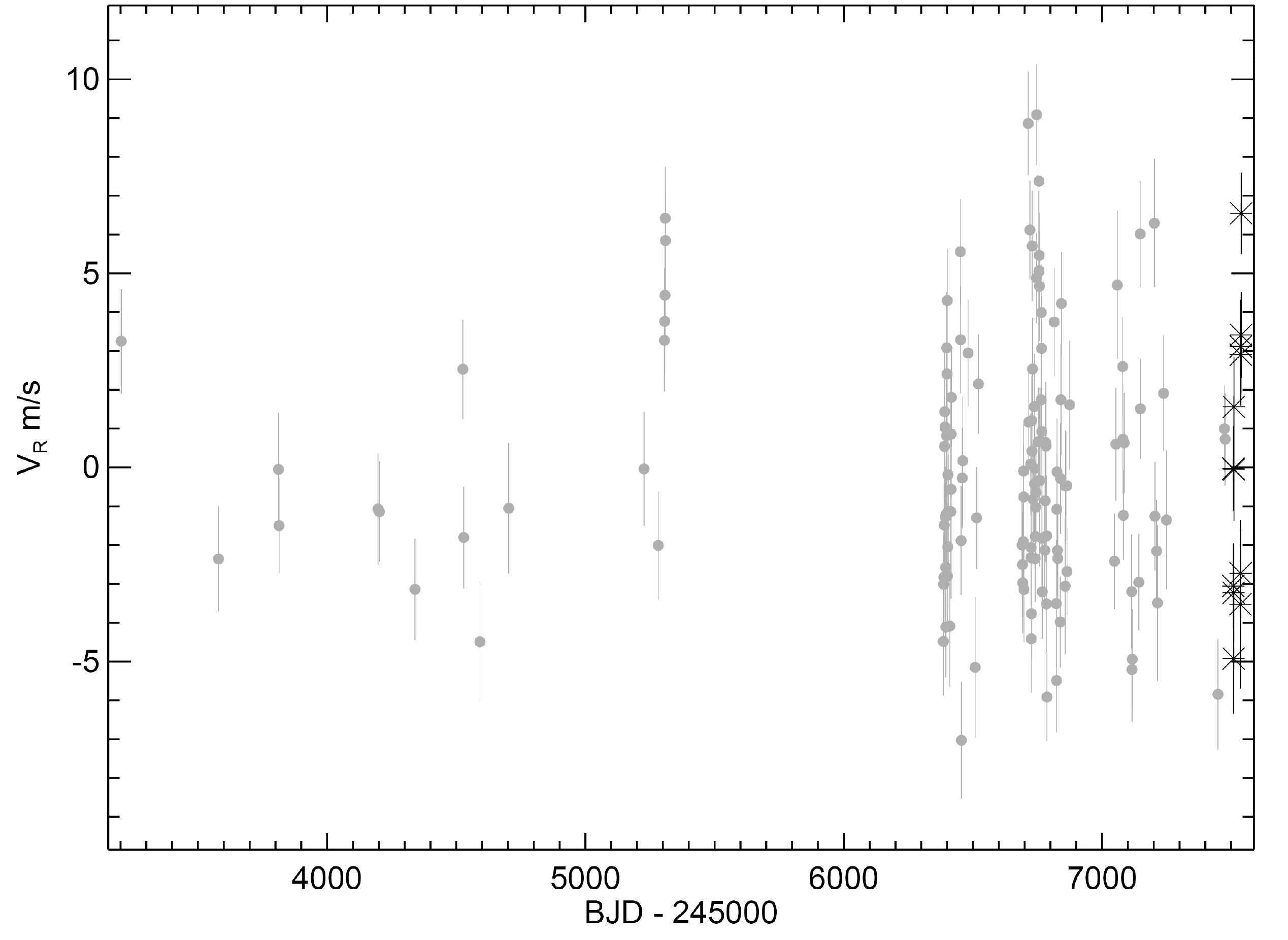}
        \caption{Radial-velocity time series. Grey dots show HARPS-S data; black asterisks show HARPS-N data.}
        \label{rv_ts0}
\end{figure}

Our 152 radial-velocity measurements have a median error of 1.33 ms$^{-1}$ which includes both photon noise, calibration, and telescope related errors. We measure a total systematic radial velocity of -25.622 Kms$^{-1}$ with a dispersion of 3.28 ms$^{-1}$. Figure~\ref{rv_ts0} shows the measured radial velocities. An F-test \citep{Zechmeister2009b} returns a negligible probability (smaller than the
0.1\%) that the internal errors explain the measured dispersion. 

To search for periodic radial-velocity signals in our time-series we follow a similar procedure to the one explained in section 3.1. We search for periodic signals using a generalised Lomb-Scargle periodogram, and if there is any significant periodicity we fit the detected signal using the RVLIN package \citep{WrightHoward2012}. We sequentially find the dominant components in the time series and remove them until  no  significant signal remains. 

\begin{figure}
        \includegraphics[width=9.0cm]{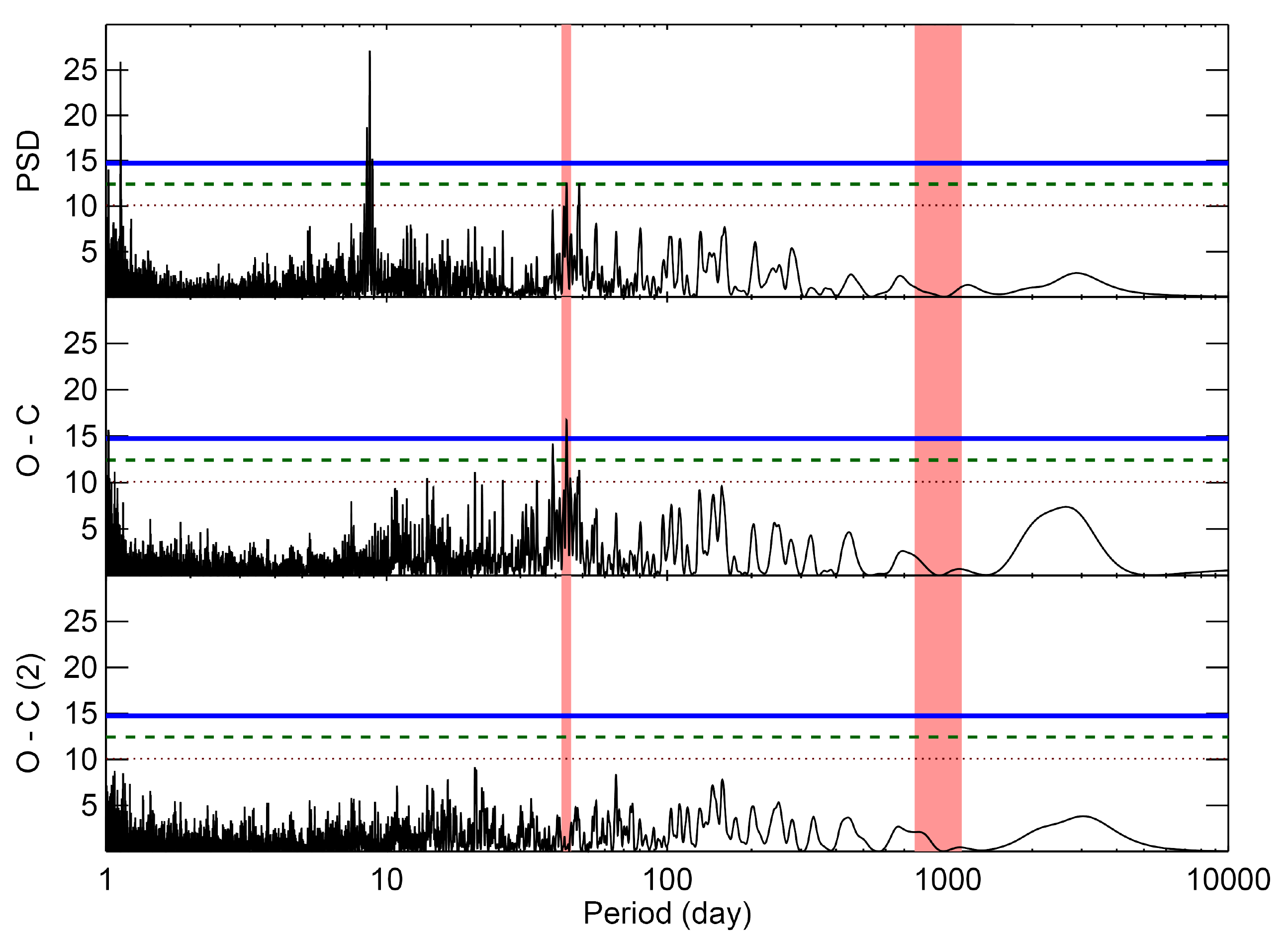}
        \caption{Periodograms of the radial velocity. The upper panel shows the raw periodogram, the middle panel the periodogram of the residuals after subtracting the 8.7 d signal, and the lower panel the periodogram of the residuals after subtracting the 43 d signal present in the middle panel. Red regions show the periods of the measured rotation and magnetic cycle. Red dotted line for a 10\%  false alarm probability; green dashed line for  1\%, and blue thick line for  0.1\%.}
        \label{rv_ts}
\end{figure}

\begin{figure}
        \includegraphics[width=9.0cm]{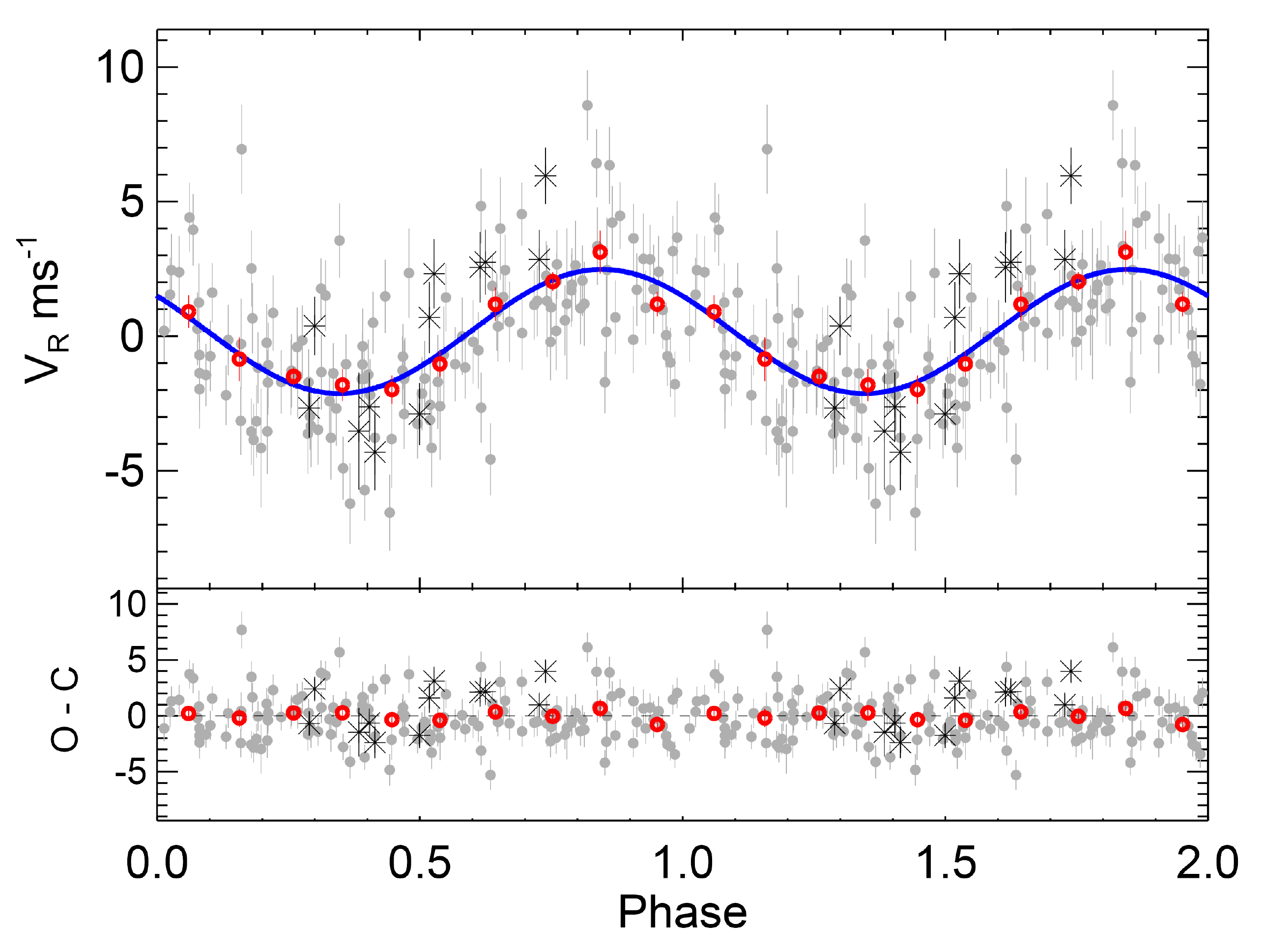}
        \includegraphics[width=9.0cm]{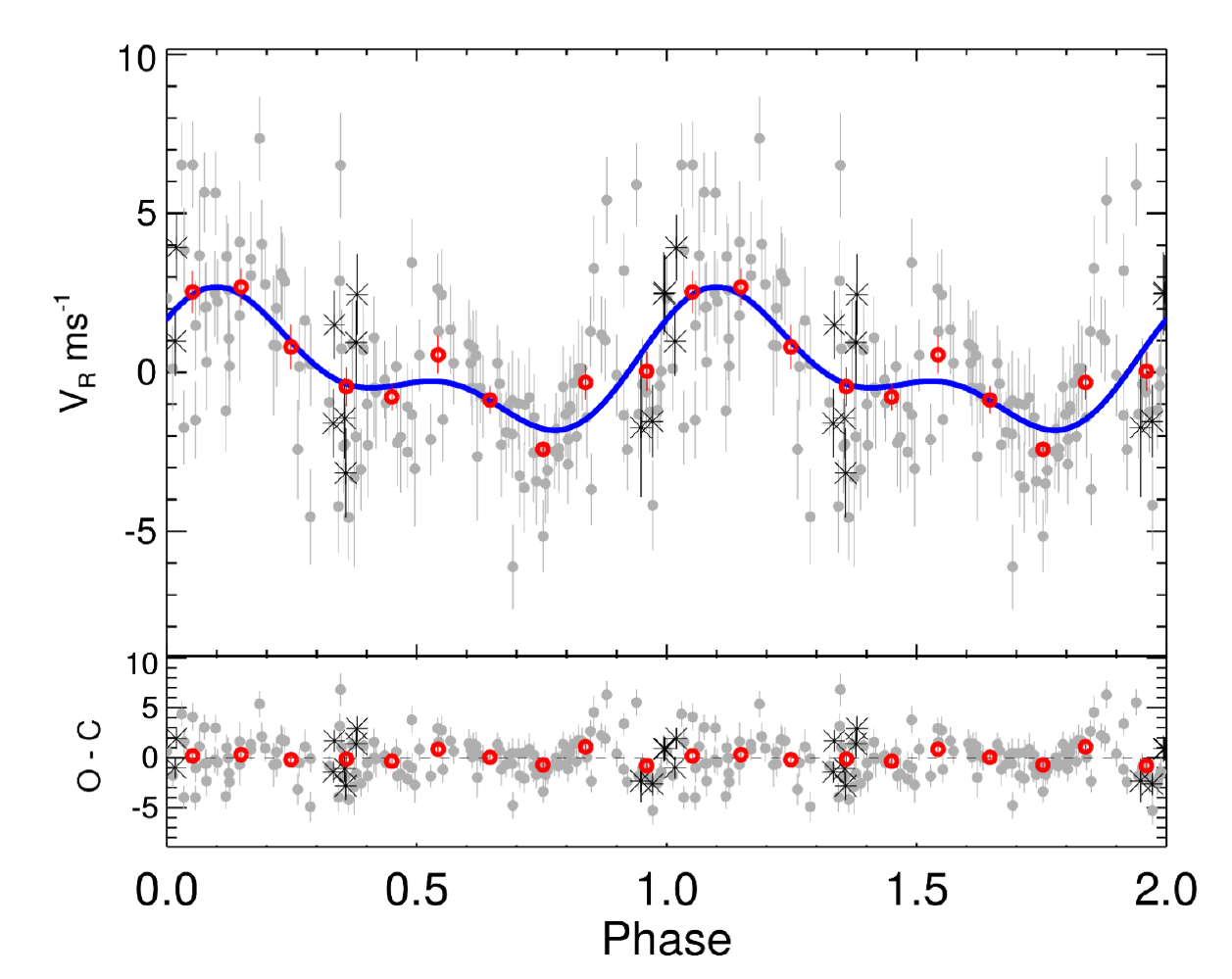}
        \caption{\textbf{Top panel:} Phase folded curve of the radial velocity using the 8.7    d period. Grey dots are the raw radial-velocity measurements after subtracting the mean value and the 43.9 d signal. \textbf{Bottom panel:} Phase folded curve of the radial velocity using the 43.9         d period using a double-harmonic sine curve. Grey dots and black asterisks are the raw radial-velocity measurements after subtracting the mean value and the 8.7 d signal.
         Red dots are the same points binned in phase with a bin size of 0.1. The error bar of a given bin is estimated using the weighted standard deviation of binned measurements divided by the square root of the number of measurements included in this bin. This estimation of the bin error bars assumes white noise, which is justified by the binning in phase and which regroups points that are uncorrelated in time.}
        \label{rv_phase}
\end{figure}

Following this procedure we identify one signal with a false alarm probability much higher than  0.1\%,  using both the bootstrap and the \citet{Cumming2004} estimates, corresponding to a period of 8.7 d with a semi-amplitude of 2.47 ms$^{-1}$ consistent with circular (Fig.~\ref{rv_ts} shows the periodogram). Removing this signal leaves  a 43.9 d signal with a semi-amplitude of 2.86 ms$^{-1}$ and an eccentricity of 0.57, with a false alarm probability better than  0.1\%. No further significant signals are found after removing these  two (Fig.~\ref{rv_ts}). Fig.~\ref{rv_phase} shows the phase folded fits of both the 8.7 d and the 43.9 d signals. 

We tested the available dataset for the three ways of calculating the radial velocity, obtaining virtually the same results in every case. Results are shown for the Gaussian + polynomial fit of the cross-correlation function. 

\subsection{Origin of the periodic radial-velocity signals}

Stellar activity can induce radial-velocity signals similar to those of Keplerian origin. The inhomogeneities in the surface of the star cause radial-velocity shifts due to the distortion of the spectral line shapes which can, in some cases, create a radial-velocity signal with a periodicity close to the stellar rotation and its first harmonic. 

For this star we have a rotation period of 45.39 $\pm$ 1.33 d, and two radial-velocity signals of 8.7 d and 43.9 d. The second signal matches almost perfectly the rotation period of the star. On the other hand we do not see in the time series of activity indicators any signal close to the 8.7 d. This is the first evidence of the stellar origin of the 43.9 d signal, and the planetary origin of the 8.7 d signal. 

\begin{figure}
        \includegraphics[width=9cm]{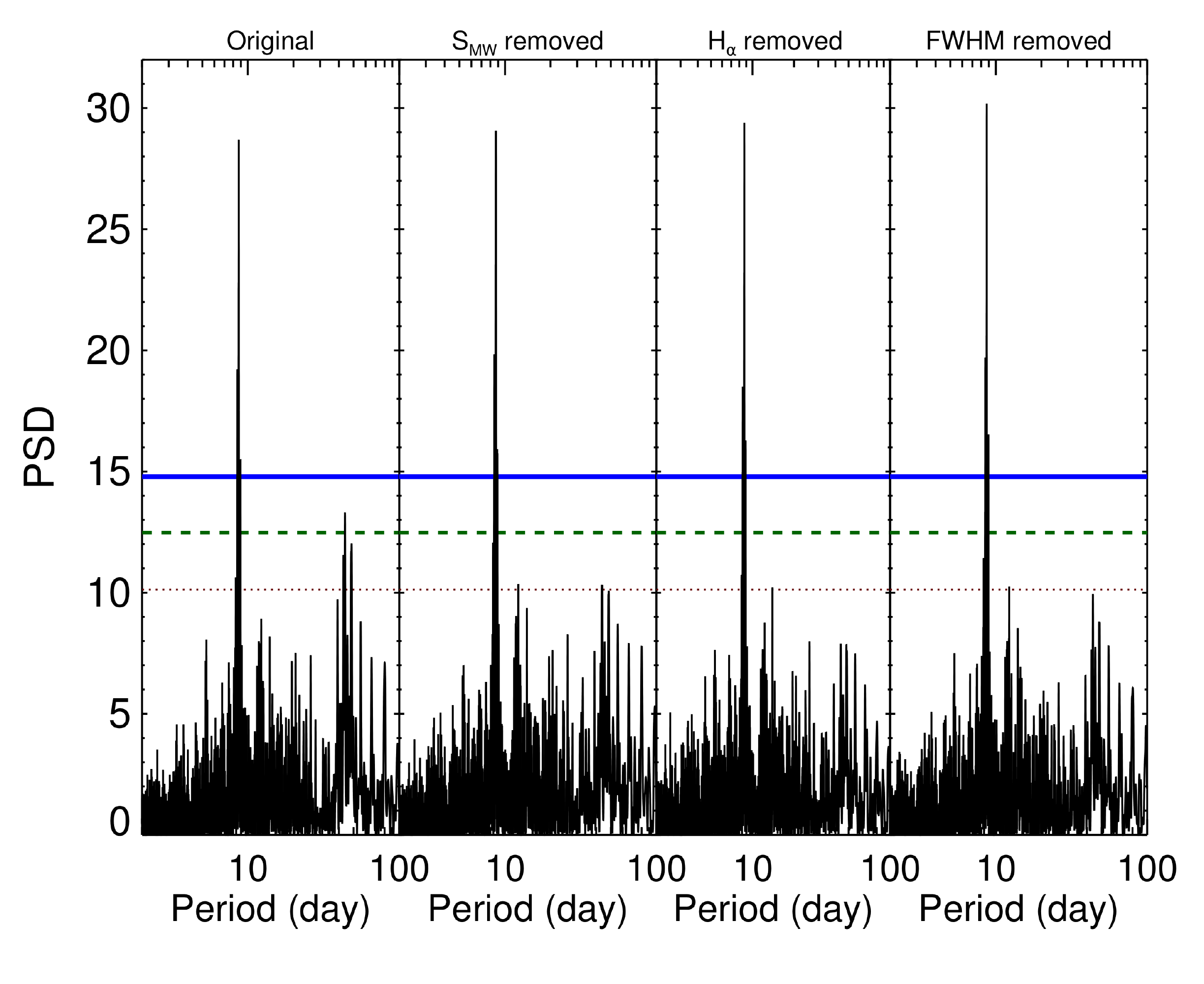}
        \caption{Periodograms for the radial velocity after removing the correlation with the different activity diagnostic tools. From left to right there is the periodogram for the original data, the periodogram after detrending against the S$_{MW}$ index, against the H$_{\alpha}$ index, and against the FWHM.}
        \label{rv_nocorr}
\end{figure}

As a second test we measured the Spearman correlation coefficient between the S$_{MW}$, the H$_{\alpha}$ index, the FWHM, and the radial velocities. We find a significant correlation between all the indexes and the raw radial velocity, which almost disappears when we isolate the 8.7 d signal, and  slightly increases when isolating the 43.9 d signal (see Table.~\ref{rv_corre}).   This constitutes a second piece of  evidence of the stellar origin of the 43.9 d signal, and of the planetary origin of the 8.7 d signal. Following this idea, we subtract the linear correlation between the radial velocity and each of the three activity diagnostic indexes. By doing this we see that the strength of the 8.7 d signal remains constant, or even  increases, while the significance of the 43.9 d is reduced in all cases (see Fig.~\ref{rv_nocorr}), even getting buried in the noise after correcting for the correlation with the H$_{\alpha}$ index. 

\begin{figure}
        \includegraphics[width=9cm]{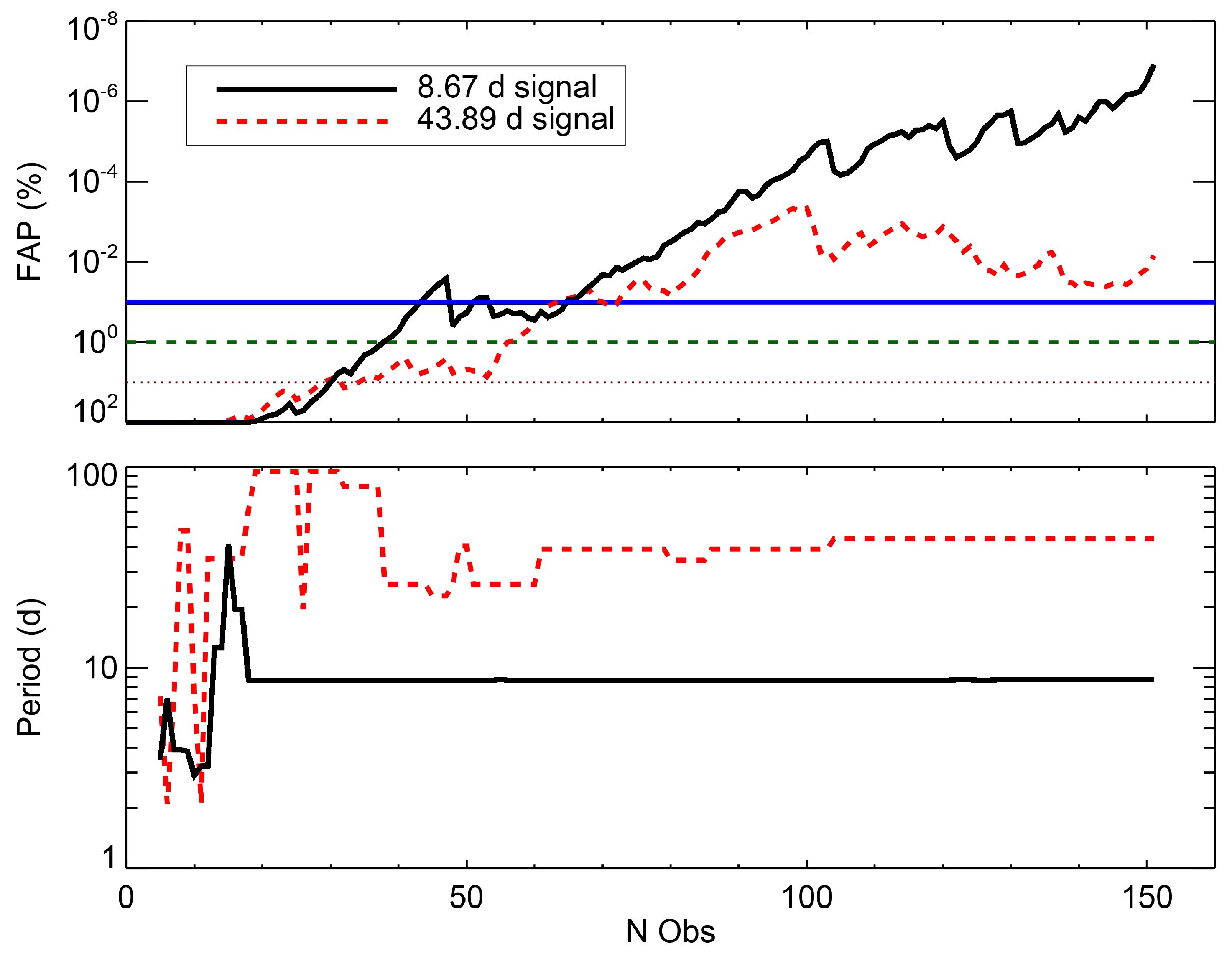}
        \caption{Evolution of the false alarm probability of the detections (upper panel) for the isolated signals, and stability of the detections (lower panel). Blue thick line shows the behaviour for the 8.7 d signals and red dashed line for the 43.9 d signal.}
        \label{rv_signals}
\end{figure}

Keplerian signals are deterministic and consistent in time. When measuring one signal, we expect to find that the significance of the detection increases steadily with the number of observations, and that the measured period is stable over time. However, in the case of an activity related signal this is not necessarily the case. As the stellar surface is not static, and the configuration of active regions may change in time, changes in the phase of the modulation and in the detected period are expected. Even the disappearance of the signal at certain seasons is possible. Fig.~\ref{rv_signals} shows the evolution of the false alarm probability of the detection of both isolated signals, as well as the measurement of the most prominent period when isolating them. The 8.7 d signal increases steadily with time, and once it becomes the most significant signal it never moves again. On the other hand, the behaviour of the 43.9 d is more erratic, losing significance during the last observations. 

Of the two significant radial-velocity signals detected in our data it seems clear that the one at 8.7 d has a planetary origin, while the one at 43.9 has a stellar activity  origin. 

The shape of the activity induced radial-velocity signal present in our data is evidently not sinusoidal. A double harmonic sinusoidal, as in the case of the activity signals, is the best fit model and the only one that does not create ghost signals after subtracting it. The rotation induced signal is not in phase with the rotation signals in the activity indicators. It appears to be  shifted by $\sim$45$^\circ$  from the signal in the S$_{MW}$ index and FWHM time series as seen in \citet{Bonfils2007} and \citet{Santos2014}. The uncertainty in the phase H$_\alpha$ time series makes it difficult to measure a reliable phase difference.

\begin {table}
\begin{center}
\caption {Activity - Radial-velocity correlations \label{tab:rv_corre}}
    \begin{tabular}{ l  l  l l l l l l l l l l } \hline
Parameter & Raw data & 8.7 d signal &  43.9 d signal &\\ \hline
S$_{MW}$ vs V$_{R}$ & 0.292 ($\textgreater$ 3$\sigma$) & 0.069 ($\textless$1$\sigma$) & 0.345 ($\textgreater$ 3$\sigma$)\\
H$_{\alpha}$ vs V$_{R}$ & 0.338 ($\textgreater$3$\sigma$) & 0.113 (1$\sigma$) & 0.321 ($\textgreater$3$\sigma$)\\
FWHM vs V$_{R}$ & 0.356 ($\textgreater$ 3$\sigma$) & 0.164 (1$\sigma$) & 0.340 ($\textgreater$ 3$\sigma$)\\
\label{rv_corre}
\end{tabular}  
\end{center}
Long-term variations of activity indicators have been subtracted. 
The parenthesis value indicates the significance of the correlation given by the bootstrapping process. 
\end {table}

Finally, an analysis of the spectral window  rules out that the peaks in the periodogram are artefacts of the time sampling alone. No features appear at 8.7 or 43.9 days even after masking the oversaturated regions of the power spectrum. Following \citet{Rajpaul2016} we tried to re-create the 8.7-day by injecting the P$_{\rm Rot}$ signal along with a second signal at P$_{\rm Rot}$/2 at 1000 randomized phase shifts with a white noise model. We were never able to generate a signal at 8.7 days, or any significant signal at periods close to 8.7 days. It seems very unlikely that any of the signals are artefacts of the sampling.

\subsection{GJ 536 b}

The analysis of the radial-velocity time series and of the activity indicators leads us to conclude that the best explanation of the observed data is the existence of a planet orbiting the star GJ 536 with a  period of 8.7 d, with a semi-amplitude of $\sim$ 2.5 ms$^{-1}$. The best solution comes from a super-Earth with a minimum mass of 5.3 M$_{\oplus}$ orbiting at 0.067 AU of its star.

\subsection*{MCMC analysis of the radial-velocity time series}

In order to quantify the uncertainties of the orbital parameters of the planet,
we perform a Bayesian analysis using the code {\sc ExoFit}~\citep{Balan2009}.
This code follows the Bayesian method described in~\citet{Gregory2005,Ford2005} and \citet{FordGregory2007}. 
A single planet can be modelled using the following formula:
\begin{equation}
        v_i = \gamma - K [\sin( \theta(t_i+\chi P) + \omega ) + e \sin \omega ]
,\end{equation}where $\gamma$ is system radial velocity; $K$ is the velocity semi-amplitude equal to
$2\pi P^{-1} (1-e^2)^{-1/2} a \sin i$; $P$ is the orbital period; $a$ is the semi-major
axis of the orbit; $e$ is the orbital eccentricity; $i$ is the inclination of the orbit;
$\omega$ is the longitude of periastron; $\chi$ is the fraction of an orbit, prior to 
the start of data taking, at which periastron occurs (thus, $\chi P$ equals the 
number of days prior to $t_i=0$ that the star was at periastron, for an orbital period 
of $P$ days); and $\theta(t_i + \chi P)$ is the angle of the star in its orbit 
relative to periastron at time $t_i$, also called the true anomaly.

To fit the previous equation to the data we need to specify the six model 
parameters, $P$, $K$, $\gamma$, $e$, $\omega$, and $\chi$.
Observed radial-velocity data, $d_i$, can be modelled by the equation $d_i = v_i + \epsilon_i + \delta$ \citep{Gregory2005}, where $v_i$ is 
the modelled radial velocity of the star and $\epsilon_i$ is the uncertainty 
component arising from accountable but unequal measurement error, which are 
assumed to be normally distributed. The term $\delta$ explains any unknown 
measurement error. Any noise component that cannot be modelled is described 
by the term $\delta$. 
The probability distribution of $\delta$ is chosen to be a Gaussian distribution 
with finite variance $s^2$. Therefore, the combination of uncertainties 
$\epsilon_i + \delta$ has a Gaussian distribution with a variance equal to 
$\sigma_i^2 + s^2$~\citep[see][for more details]{Balan2009}.

The parameter estimation in the Bayesian analysis needs a choice of priors. We choose the priors following the studies by \citet{FordGregory2007, Balan2009}.The mathematical form of the prior is given in Table 1 and/or 4 of \citet{Balan2009}. In Table~\ref{mcmc_par}, we provide the parameter boundaries explored in the MCMC Bayesian analysis. {\sc ExoFit} performs 100 chains of 10000 iterations each resulting in a final chain of 19600 sets of global-fit parameters.

\begin{figure*}
        \includegraphics[width=18cm]{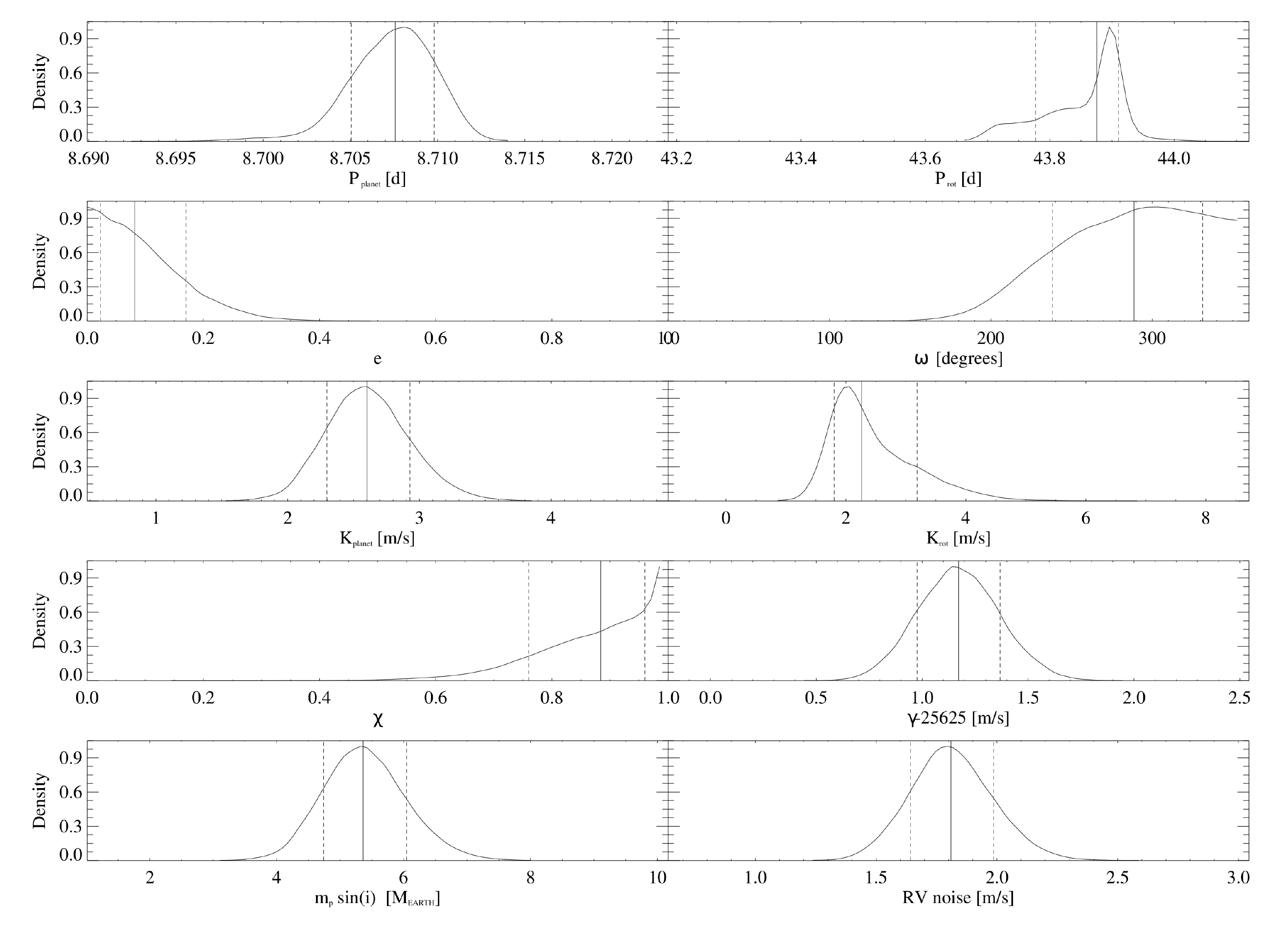}
\caption{Posterior distribution of model parameters including the activity 
signal associated with the rotation period and the orbital parameters of 
the planet companion of the M-dwarf star GJ~536. The vertical dashed line shows the median
value of the distribution and the dotted lines the 1-$\sigma$ values.}
\label{mcmc}
\end{figure*}

We want to simultaneously model the stellar rotation and planetary signals. For that 
we use the {\sc ExoFit} to model two RV signals and for the rotation signal we also 
leave the eccentricity as a free parameter. The posterior distribution of the eccentricity parameter 
for the rotation signal (not shown in Fig.~\ref{mcmc}) gives 
a value of $0.47 \pm 0.26$.
In Fig.~\ref{mcmc} we depict the posterior distribution of the model parameters; the 
six fitted parameters; the semi-amplitude velocity, $K_{\rm rot}$, and the period, 
$P_{\rm rot}$, of the rotation signal; the derived mass of the planet, $m_p \sin i$;
and the RV noise given by the $s$ parameter.
Most of the parameters show symmetric density profiles except for the eccentricity, 
$e$; the longitude of periastron, $\omega$; and the fraction $\chi$ of the orbit at which 
the periastron occurs. We note that the density profile of the rotation period displays 
a tail towards slightly lower values although the rotation period is well defined.

In Table~\ref{mcmc_par} we show the final parameters and uncertainties obtained 
with the MCMC Bayesian analysis with the code {\sc ExoFit}.

\begin{table*}
\begin{center}
\caption {MCMC parameters and uncertainties\label{mcmc_par}}
\begin{tabular}{l c c c c}
\hline
Parameter & Value & Upper error & Lower error & Prior\\
\hline
$P_{\rm planet}$ [d]             & 8.7076 & $+$0.0022 & $-$0.0025 & 8.3 - 9.0 \\
$\gamma$ [ms$^{-1}-25625$]       & 1.17   & $+$0.20   & $-$0.20   & $-$5.0 - $+$5.0 \\
$e$                              & 0.08   & $+$0.09   & $-$0.06  & 0.0 - 0.99 \\
$\omega$ [deg]                   & 288.7  & $+$42.5   & $-$50.6  & 0.0 - 360.0 \\
$\chi$                           & 0.88   & $+$0.08   & $-$0.12  & 0.0 - 0.99 \\
$K_{\rm planet}$ [ms$^{-1}$]     & 2.60   & $+$0.33   & $-$0.30  & 0.0 - 5.0 \\
$a$ [AU]                         & 0.066610 & $+$0.000011 & $-$0.000013  & -- \\
$m_p \sin i$ [M$_{\rm Earth}$]   & 5.36   & $+$0.69   & $-$0.62  & -- \\
$P_{\rm rot}$ [d]                & 43.88  & $+$0.03   & $-$0.10  & 42.5 - 45.0 \\
$K_{\rm rot}$ [ms$^{-1}$]        & 2.26   & $+$0.92   & $-$0.46  & 0.0 - 7.0 \\
RV noise [ms$^{-1}$]             & 1.81   & $+$0.18   & $-$0.17  & 0.0 - 5.0 \\
\hline
\end{tabular}  
\end{center}
\end{table*}

\section{Discussion}

We detect the presence of a planet with a semi-amplitude of 2.60 m s$^{-1}$, which -- given the stellar mass of 0.52 M$_{\odot}$ -- converts to m sin $i$ of 5.36 M$_{\oplus}$, orbiting with a period of 8.7 d around GJ 536, an M-type star of 0.52 M$_{\odot}$ with a rotation period of 43.9 d that shows an additional activity signal compatible with an  activity cycle shorter than 3 yr.

The planet is a small super-Earth with an equilibrium temperature 344 K for a Bond albedo A = 0.75 and  487 K for A=0. Following \citet{Kasting1993} and \citet{Selsis2007}, we perform a simple estimation of the habitable zone (HZ) of this star. The HZ would go from 0.2048 to 0.3975 AU in the narrowest case (cloud free model), and from 0.1044 to 0.5470 AU in the broadest case (fully clouded model). This corresponds to orbital periods ranging from 46 to 126 days in the narrowest case, and from 17 to 204 days in the broadest one. 

\begin{figure*}
        \includegraphics[width=18cm]{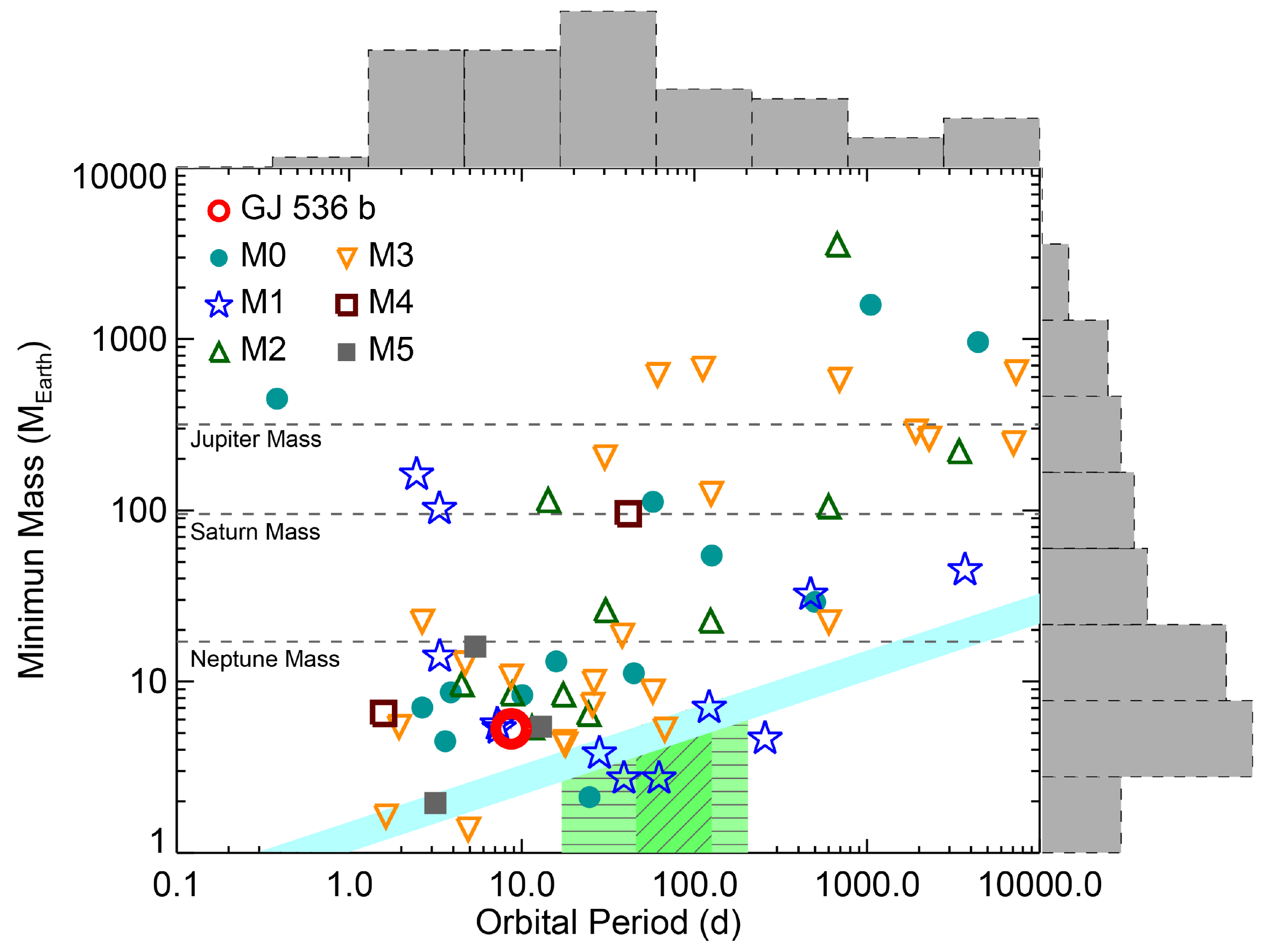}
        \caption{Minimum mass vs orbital period for the known planets around M-dwarf stars. The red empty dot shows the position of GJ 536 b.  Horizontal dashed lines show the mass of the solar system planets for comparison. On the edges of the figure the distribution for each parameter is shown. The cyan shape shows the region where a second planet could exist around GJ 536 (upper limit of the band) and could be detectable given the typical uncertainty of the measurements (lower limit of the band). The green shapes show the habitable region around the star GJ 536. Horizontal lines show the broad scenario, while inclined lines show the narrow one.}
        \label{period_mass}
\end{figure*}

GJ 536 b is  in the lower part of the mass vs period diagram of  known planets around M-dwarf stars (Fig.~\ref{period_mass}). The planet is too close to the star to be considered habitable. For this star the orbital periods of the habitable zone would be from $\sim$20 days to $\sim$ 40 days.

GJ 536 is a quiet early M dwarf, with a rotation period at the upper end of the stars of its kind \citep{Newton2016,Masca2016a}. Its rotation induced radial-velocity signal has a semi-amplitude of 2.26 ms$^{-1}$ and seems to be stable enough to allow for a clean enough periodogram and to be correctly characterized.  The phase of the rotation induced signal seems to be advanced by $\sim$45$^\circ$ with respect to the signals in the S$_{MW}$ index and FWHM time series. There is a hint of an  activity cycle shorter than 3 yr, which would put it at the lower end of the stars of its kind \citep{Masca2016a}, and whose amplitude is so small that would need further follow-up to be properly characterized. The radial-velocity signal induced by this cycle  at this point is beyond our detection capabilities.

Given the rms of the residuals there is still room for the detection of more planets in this system, especially at orbital periods longer than the rotation period. Fig.~\ref{period_mass} shows the upper limits to the mass of those hypothetical companions. The stability of its rotation signals and the low amplitude of the radial-velocity signals with a magnetic origin makes this star a good candidate  to search for longer period planets of moderate mass. A rough estimate of the detection limits tells us there is still room for Earth-like planets ($\sim$ 1 M$_{\oplus}$) at orbits smaller than 10, super-Earths ($\textless$ 10 M$_{\oplus}$) at orbits going from 10 to 400 days, and even for a Neptune-mass planet ($\textless$ 20 M$_{\oplus}$)   at periods longer than $\sim$3 yr. Giant planets, on the other hand, are discarded except for those with extremely long orbital periods. The time-span of the observations and the RMS of the residuals completely rules out the presence of any planet bigger than twice the mass of Neptune with an orbital period shorter than $\sim$ 20 years.

\section{Conclusions}

We have analysed 152 high-resolution spectra and 359 photometric observations to study the  planetary companions around the M-dwarf star GJ 536 and its stellar activity. 
We detected two significant radial-velocity signals at periods of 8.7 and 43.8 days, respectively.

From the available photometric and spectroscopic information  we conclude that the 8.7 d signal is caused by a  5.3 M$_{\oplus}$ planet with semi-major axis of 0.067 AU and  equilibrium temperature lower than 500K. The short period of the planet makes it a potential transiting candidate. Detecting the transits would give a new constraining point to the mass-radius diagram. 

The second radial-velocity signal of period 43.8 d and semi-amplitude of 1.6 ms$^{-1}$ is a magnetic activity induced signal related to the rotation of the star.  We also found a magnetic cycle shorter than 3 yr which would place this star among those with the shortest reported  magnetic cycles.

We have studied and set limits to the presence of other planetary companions taking into account the rms of the residuals after fitting both the planet and the rotation induced signal. The system still has room for other low-mass companions, but planets more massive than Neptune are discarded except at extremely long orbital periods beyond the habitable zone of the star.

\section*{Acknowledgements}
This work has been financed by the Spanish Ministry project MINECO AYA2014-56359-P. J.I.G.H. acknowledges financial support from the Spanish MINECO under the 2013 Ram\'on y Cajal program MINECO RYC-2013-14875.  X.B., X.D., T.F., and F.M. acknowledge the support of the French Agence Nationale de la Recherche (ANR) under the programme ANR-12-BS05- 0012 Exo-atmos. X.B. and A.W. acknowledge funding from the European Research Council under the ERC Grant Agreement No. 337591-ExTrA. This work was supported by Funda\c{c}\~ao para a Ci\^encia e a Tecnologia (FCT) within projects reference PTDC/FIS-AST/1526/2014 (POCI-01-0145-FEDER-016886) and UID/FIS/04434/2013 (POCI-01-0145-FEDER-007672). N.C.S. acknowledges support   through Investigador FCT contract of reference IF/00169/2012, and POPH/FSE (EC) by FEDER funding through the program ``Programa Operacional de Factores de Competitividade - COMPETE''. This work is based on data obtained via the  HARPS public database at the European Southern Observatory (ESO). This research has made extensive use of the SIMBAD database, operated at CDS, Strasbourg, France, and NASA's Astrophysics Data System. We are grateful to all the observers of the following ESO projects, whose data we are using: 072.C-0488, 085.C-0019, 183.C-0972, and 191.C-087.

%
%
\bibliography{RHK_ref}

\onecolumn
\begin{appendix}
\section{Full Dataset} \label{append_a}

\begin{longtable}{@{\extracolsep{\fill}}llccccccccccc@{}}
\caption {Full available dataset. Radial velocities are given in the Barycentric Reference Frame after subtracting the secular acceleration. Radial-velocity uncertainties include photon noise, calibration and telescope related uncertainties. \label{tab_a1}}\\
\endfirsthead
\multicolumn{9}{c}{\tablename\ \thetable\ -- \textit{Continued from previous page}} \\
\hline
\endhead
\hline \multicolumn{9}{r}{\textit{Continued on next page}} \\
\endfoot
\hline
\endlastfoot
\hline

BJD - 2450000 & V$_{r}$  & $\sigma$ V$_{r}$ &  FWHM  & BIS Span  & S$_{MW}$ index & $\sigma$ S$_{MW}$ & H$_{\alpha}$ index & $\sigma$ H$_{\alpha}$ index & Flag\\ 
 (d) & (ms$^{-1}$) & (ms$^{-1}$) & (Kms$^{-1}$) & (ms$^{-1}$) \\ \hline
3202.5590 & -25616.5610 & 1.3488 & 4444.8091 & -5.6623 & 1.0772 & 0.0077 & 0.4960 & 0.0005\\
3579.4972 & -25622.4158 & 1.3573 & 4448.4550 & -5.0540 & 1.2490 & 0.0073 & 0.5058 & 0.0005\\
3811.8370 & -25620.2605 & 1.4564 & 4438.6871 & -9.9863 & 1.1320 & 0.0078 & 0.5043 & 0.0006\\
3813.8047 & -25621.7093 & 1.2228 & 4433.0342 & -8.1469 & 1.1526 & 0.0053 & 0.5047 & 0.0004\\
4196.7394 & -25621.5343 & 1.4416 & 4443.0970 & -6.2410 & 1.0422 & 0.0074 & 0.5001 & 0.0006\\
4202.7156 & -25621.6073 & 1.2951 & 4425.1466 & -6.9079 & 1.0538 & 0.0059 & 0.5039 & 0.0005\\
4340.4836 & -25623.6979 & 1.3024 & 4429.4257 & -7.9120 & 1.1163 & 0.0070 & 0.5076 & 0.0005\\
4525.8756 & -25618.1517 & 1.2817 & 4433.4956 & -11.6982 & 1.1585 & 0.0063 & 0.5069 & 0.0005\\
4528.8393 & -25622.4852 & 1.3126 & 4433.8369 & -5.2346 & 1.1304 & 0.0064 & 0.5044 & 0.0005\\
4591.7914 & -25625.2134 & 1.5565 & 4442.0880 & -7.2876 & 0.9757 & 0.0090 & 0.4940 & 0.0007\\
4703.4993 & -25621.8483 & 1.6802 & 4433.6780 & -10.1303 & 1.1269 & 0.0107 & 0.5032 & 0.0008\\
5226.8854 & -25621.1796 & 1.4632 & 4446.4284 & -10.0405 & 1.2068 & 0.0081 & 0.4996 & 0.0006\\
5281.7491 & -25623.1861 & 1.3854 & 4428.1838 & -8.6441 & 1.0451 & 0.0068 & 0.4889 & 0.0006\\
5305.7265 & -25617.9212 & 1.3201 & 4427.4995 & -9.9302 & 1.0898 & 0.0084 & 0.5030 & 0.0005\\
5306.7140 & -25617.4308 & 1.3665 & 4430.6156 & -8.9673 & 1.1605 & 0.0092 & 0.5063 & 0.0006\\
5307.7196 & -25616.7602 & 1.2588 & 4436.3939 & -10.4465 & 1.2272 & 0.0087 & 0.5109 & 0.0004\\
5308.7013 & -25614.7745 & 1.3191 & 4446.0321 & -8.0523 & 1.1629 & 0.0088 & 0.5039 & 0.0005\\
5309.6925 & -25615.3506 & 1.3031 & 4442.9390 & -8.8002 & 1.2196 & 0.0092 & 0.5080 & 0.0005\\
6385.6469 & -25626.3858 & 1.3821 & 4428.1616 & -3.0622 & 0.8991 & 0.0101 & 0.4978 & 0.0007\\
6386.7448 & -25624.9118 & 1.2594 & 4425.8662 & -9.9115 & 0.8084 & 0.0078 & 0.4935 & 0.0007\\
6387.7815 & -25624.7334 & 1.1527 & 4424.2671 & -8.2783 & 0.8489 & 0.0073 & 0.4952 & 0.0006\\
6388.7254 & -25623.3896 & 1.3647 & 4416.3803 & -7.3567 & 0.8284 & 0.0087 & 0.4964 & 0.0007\\
6389.7264 & -25621.3622 & 1.3006 & 4421.5546 & -11.3118 & 0.8655 & 0.0085 & 0.5048 & 0.0007\\
6390.7371 & -25620.4734 & 1.6898 & 4419.3411 & -9.8647 & 0.8066 & 0.0105 & 0.4943 & 0.0008\\
6391.7497 & -25620.8694 & 1.3286 & 4424.2490 & -11.7541 & 0.9001 & 0.0085 & 0.5003 & 0.0007\\
6393.7913 & -25623.1815 & 1.2910 & 4415.1907 & -7.5922 & 0.8531 & 0.0089 & 0.4978 & 0.0007\\
6394.7750 & -25624.4828 & 1.2851 & 4429.5659 & -9.1806 & 0.8915 & 0.0088 & 0.4949 & 0.0007\\
6395.7000 & -25626.0196 & 1.2787 & 4417.1293 & -9.7255 & 0.9155 & 0.0089 & 0.4957 & 0.0007\\
6396.7103 & -25623.1147 & 1.4092 & 4424.7395 & -12.7127 & 0.9218 & 0.0094 & 0.4968 & 0.0008\\
6397.6863 & -25621.0933 & 1.3000 & 4423.8606 & -9.5686 & 0.9836 & 0.0095 & 0.5064 & 0.0007\\
6398.6799 & -25618.8329 & 1.4313 & 4426.1457 & -7.6371 & 0.9339 & 0.0099 & 0.4992 & 0.0008\\
6399.6958 & -25619.5072 & 1.3309 & 4440.7943 & -9.3043 & 1.0242 & 0.0097 & 0.5024 & 0.0008\\
6400.6899 & -25617.6148 & 1.3276 & 4426.9125 & -10.7879 & 0.9848 & 0.0094 & 0.5022 & 0.0008\\
6401.6532 & -25624.7094 & 1.2229 & 4430.4209 & -8.8857 & 1.0296 & 0.0087 & 0.5028 & 0.0007\\
6402.6436 & -25623.9620 & 1.5248 & 4430.2528 & -9.1200 & 1.0116 & 0.0108 & 0.5027 & 0.0008\\
6403.6245 & -25622.1027 & 1.3310 & 4443.0225 & -9.7811 & 1.0437 & 0.0098 & 0.5032 & 0.0007\\

 6404.6425 & -25623.0456 & 1.4532 & 4434.0050 & -10.1067 & 1.0945 & 0.0109 & 0.5106 & 0.0008\\
6410.6262 & -25626.0065 & 1.5706 & 4448.4592 & -8.4057 & 0.9966 & 0.0115 & 0.5039 & 0.0008\\
6414.6393 & -25623.0607 & 1.4853 & 4439.0697 & -6.4884 & 1.0073 & 0.0104 & 0.5072 & 0.0008\\
6415.5922 & -25622.4846 & 2.8093 & 4434.8839 & -17.9962 & 0.8264 & 0.0184 & 0.5053 & 0.0014\\
6415.7332 & -25621.0645 & 1.6543 & 4445.0213 & -9.2692 & 1.0473 & 0.0124 & 0.5114 & 0.0009\\
6416.6954 & -25620.1184 & 1.2661 & 4431.5240 & -4.3249 & 0.9507 & 0.0089 & 0.5034 & 0.0007\\
6451.5800 & -25616.3891 & 1.3405 & 4433.3694 & -16.8458 & 1.0419 & 0.0096 & 0.5063 & 0.0007\\
6452.5545 & -25618.6586 & 1.3803 & 4434.6138 & -8.7892 & 1.0683 & 0.0081 & 0.5099 & 0.0007\\
6454.5556 & -25623.8325 & 1.3937 & 4431.7379 & -9.5482 & 1.0701 & 0.0103 & 0.5075 & 0.0008\\
6455.5374 & -25628.9792 & 1.4965 & 4422.2962 & -7.9232 & 0.9761 & 0.0106 & 0.5020 & 0.0008\\
6458.5877 & -25622.2235 & 1.2855 & 4423.7546 & -11.6472 & 1.0516 & 0.0093 & 0.5101 & 0.0007\\
6460.5668 & -25621.7813 & 1.6473 & 4431.8028 & -10.6556 & 0.9902 & 0.0115 & 0.5048 & 0.0008\\
6481.4839 & -25619.0212 & 1.3733 & 4419.7151 & -8.8134 & 0.9295 & 0.0089 & 0.5034 & 0.0007\\
6508.4718 & -25627.1324 & 1.8094 & 4441.7103 & -10.4168 & 0.8926 & 0.0119 & 0.5119 & 0.0010\\
 6514.4694 & -25623.2844 & 1.3027 & 4424.7201 & -6.1677 & 0.8424 & 0.0087 & 0.5033 & 0.0007\\

 6521.4589 & -25619.8433 & 1.2696 & 4431.0487 & -10.7298 & 0.8505 & 0.0081 & 0.5007 & 0.0007\\
 6690.8780 & -25624.1045 & 1.1926 & 4422.6617 & -9.2806 & 0.9930 & 0.0082 & 0.5063 & 0.0006\\
 6691.8339 & -25624.6074 & 1.3498 & 4426.5396 & -8.4985 & 1.1321 & 0.0107 & 0.5168 & 0.0007\\
 6692.8139 & -25625.0787 & 1.2900 & 4431.9081 & -10.8969 & 1.0794 & 0.0099 & 0.5048 & 0.0006\\
 6694.8640 & -25624.0179 & 1.1386 & 4433.0913 & -9.0442 & 1.0415 & 0.0081 & 0.5084 & 0.0006\\
6695.8790 & -25622.2008 & 1.1906 & 4429.7997 & -8.1605 & 1.0384 & 0.0085 & 0.5108 & 0.0006\\
6696.8539 & -25622.8668 & 1.3253 & 4428.3856 & -10.0882 & 1.2174 & 0.0107 & 0.5241 & 0.0007\\
6697.7981 & -25625.2516 & 1.3466 & 4425.1099 & -6.5195 & 1.0566 & 0.0105 & 0.5067 & 0.0007\\
6712.8127 & -25613.5661 & 1.3395 & 4437.9343 & -10.5794 & 1.5124 & 0.0112 & 0.5544 & 0.0008 & Rejected \\
6713.8033 & -25613.2583 & 1.3276 & 4450.3946 & -9.6570 & 1.2983 & 0.0109 & 0.5214 & 0.0007\\
6715.7953 & -25620.9555 & 1.3577 & 4436.1206 & -6.6456 & 1.2225 & 0.0106 & 0.5213 & 0.0007\\
6720.8502 & -25616.0066 & 1.2598 & 4432.3554 & -9.5264 & 1.1898 & 0.0094 & 0.5123 & 0.0006\\
6723.8540 & -25622.0386 & 1.2931 & 4439.7320 & -11.2245 & 1.2372 & 0.0103 & 0.5198 & 0.0007\\
6724.7853 & -25624.4497 & 1.2257 & 4440.9264 & -10.5082 & 1.2255 & 0.0097 & 0.5205 & 0.0006\\
6725.7743 & -25624.1936 & 1.2975 & 4439.1990 & -10.9028 & 1.1683 & 0.0103 & 0.5152 & 0.0007\\
6725.8844 & -25626.5410 & 1.3803 & 4440.6907 & -7.2889 & 1.1309 & 0.0105 & 0.5138 & 0.0007\\
6726.7959 & -25625.8960 & 1.1504 & 4442.3973 & -7.2719 & 1.0898 & 0.0084 & 0.5101 & 0.0006\\
6727.8296 & -25620.9225 & 1.1367 & 4430.0904 & -10.2483 & 1.0773 & 0.0079 & 0.5083 & 0.0006\\
6728.8039 & -25621.7110 & 1.1250 & 4434.7157 & -8.6990 & 1.0786 & 0.0080 & 0.5062 & 0.0005\\
6729.7718 & -25616.4237 & 1.4276 & 4437.5013 & -12.1011 & 1.0759 & 0.0085 & 0.5096 & 0.0006\\
6730.8216 & -25619.5981 & 1.3086 & 4433.6524 & -10.6152 & 1.0424 & 0.0095 & 0.5111 & 0.0007\\
6732.7980 & -25622.9502 & 1.4351 & 4421.7672 & -4.3220 & 1.0392 & 0.0101 & 0.5106 & 0.0008\\
6737.8572 & -25620.5708 & 1.3554 & 4428.3888 & -5.4130 & 1.0060 & 0.0093 & 0.5102 & 0.0007\\
6738.8726 & -25622.5574 & 1.2338 & 4431.8640 & -7.8770 & 0.9898 & 0.0086 & 0.5047 & 0.0007\\
6739.8058 & -25622.1678 & 1.1512 & 4425.2586 & -9.5216 & 1.0135 & 0.0078 & 0.5075 & 0.0006\\
6740.8311 & -25624.4843 & 1.0987 & 4428.1852 & -6.4714 & 0.9862 & 0.0073 & 0.5061 & 0.0005\\
6741.7462 & -25623.9200 & 1.1678 & 4429.0187 & -9.0739 & 1.0092 & 0.0081 & 0.5038 & 0.0006\\
6742.8207 & -25623.1636 & 1.1028 & 4426.1406 & -9.8389 & 1.0343 & 0.0076 & 0.5094 & 0.0005\\
6743.7632 & -25622.7840 & 1.2011 & 4432.8256 & -6.7370 & 1.0186 & 0.0084 & 0.5069 & 0.0006\\
6745.7321 & -25617.2597 & 1.1588 & 4423.0781 & -11.4489 & 1.0638 & 0.0081 & 0.5124 & 0.0007\\
6746.8203 & -25613.0541 & 1.2944 & 4428.6487 & -8.3582 & 1.0296 & 0.0090 & 0.5158 & 0.0007\\
6752.8315 & -25621.4802 & 1.3848 & 4432.6911 & -7.1023 & 1.0633 & 0.0106 & 0.5210 & 0.0008\\
6754.8603 & -25617.1072 & 2.0773 & 4434.8424 & -14.7497 & 1.1186 & 0.0161 & 0.5213 & 0.0011\\
6755.8430 & -25617.0817 & 2.0606 & 4444.2810 & -9.2983 & 1.0480 & 0.0157 & 0.5182 & 0.0011\\
6755.8530 & -25614.7730 & 1.9338 & 4449.5330 & -8.2994 & 1.1561 & 0.0154 & 0.5220 & 0.0010\\
6756.8521 & -25616.6832 & 1.0939 & 4440.9586 & -9.8160 & 1.1383 & 0.0085 & 0.5169 & 0.0005\\
6757.8085 & -25617.4805 & 1.4014 & 4441.8165 & -9.7412 & 1.1327 & 0.0112 & 0.5181 & 0.0007\\
6758.8266 & -25622.4824 & 2.2098 & 4438.0807 & -15.1975 & 1.0999 & 0.0174 & 0.5266 & 0.0012\\
6759.8277 & -25621.4838 & 1.3814 & 4438.3218 & -7.9763 & 1.1324 & 0.0111 & 0.5192 & 0.0007\\
6760.8142 & -25620.2626 & 1.3119 & 4448.0559 & -8.2829 & 1.5703 & 0.0125 & 0.5642 & 0.0007 & Rejected \\
6763.7243 & -25620.4110 & 1.0500 & 4442.6357 & -10.9225 & 1.1538 & 0.0073 & 0.5130 & 0.0005\\
6764.7765 & -25618.1639 & 1.2340 & 4444.4456 & -9.9255 & 1.1184 & 0.0092 & 0.5128 & 0.0006\\
6765.7208 & -25619.0905 & 1.3628 & 4437.2341 & -6.5105 & 1.1466 & 0.0078 & 0.5131 & 0.0006\\
6766.7265 & -25621.2292 & 1.0945 & 4439.1904 & -9.8447 & 1.2270 & 0.0081 & 0.5229 & 0.0005\\
6767.6534 & -25623.9784 & 1.5745 & 4433.4716 & -10.2058 & 1.1048 & 0.0115 & 0.5113 & 0.0008\\
6768.6678 & -25625.3559 & 1.2107 & 4429.1243 & -7.0854 & 1.0659 & 0.0087 & 0.5089 & 0.0006\\
6778.6271 & -25624.2948 & 1.3732 & 4424.6235 & -8.2880 & 0.9981 & 0.0077 & 0.5003 & 0.0005\\
6779.7560 & -25623.0220 & 1.5571 & 4433.6733 & -7.6866 & 0.9779 & 0.0094 & 0.5036 & 0.0007\\
6781.6011 & -25621.5230 & 1.5651 & 4434.4575 & -9.5425 & 0.9537 & 0.0096 & 0.5022 & 0.0007\\
6782.6156 & -25621.6172 & 1.3793 & 4433.3780 & -4.7485 & 1.1337 & 0.0085 & 0.5159 & 0.0005\\
6784.6137 & -25625.6796 & 1.2493 & 4430.6030 & -6.3606 & 0.9865 & 0.0089 & 0.5079 & 0.0006\\
6785.5546 & -25623.9276 & 1.5364 & 4411.9798 & -7.3440 & 0.9995 & 0.0118 & 0.5097 & 0.0008\\
6786.6679 & -25628.0817 & 1.1219 & 4420.8351 & -8.7321 & 0.9403 & 0.0073 & 0.5038 & 0.0005\\
6814.7183 & -25618.4375 & 1.3822 & 4442.1718 & -6.9710 & 1.0105 & 0.0109 & 0.5029 & 0.0007\\
6822.5823 & -25625.6969 & 1.6325 & 4427.8930 & -14.3619 & 1.0863 & 0.0119 & 0.5063 & 0.0009\\
6823.5834 & -25627.6799 & 1.3314 & 4432.6413 & -9.5886 & 1.0479 & 0.0097 & 0.5004 & 0.0007\\
6824.5777 & -25623.2701 & 1.4221 & 4430.6359 & -9.1015 & 1.0257 & 0.0097 & 0.4996 & 0.0007\\
6825.6520 & -25622.3060 & 1.3593 & 4435.5327 & -7.2742 & 1.0811 & 0.0105 & 0.5055 & 0.0007\\
6826.5764 & -25621.9304 & 1.2330 & 4428.6392 & -4.9112 & 1.3187 & 0.0094 & 0.5315 & 0.0007 & Rejected \\
6827.5754 & -25624.3321 & 1.1535 & 4432.4085 & -6.5820 & 1.0300 & 0.0082 & 0.5024 & 0.0006\\
6828.6006 & -25624.5384 & 1.1901 & 4429.2042 & -10.7163 & 1.1152 & 0.0086 & 0.5094 & 0.0006\\
6838.5568 & -25626.1804 & 1.1653 & 4430.8116 & -10.7542 & 1.0408 & 0.0083 & 0.5163 & 0.0006\\
6839.5704 & -25622.4896 & 1.4164 & 4425.8922 & -8.3993 & 0.9890 & 0.0105 & 0.5122 & 0.0006\\
6840.5286 & -25620.4583 & 1.4235 & 4421.9119 & -7.9951 & 0.9953 & 0.0097 & 0.5107 & 0.0007\\

6841.6035 & -25614.8073 & 2.5202 & 4438.8469 & -2.9369 & 1.1631 & 0.0220 & 0.5331 & 0.0013 & Rejected \\
6842.4896 & -25617.9822 & 1.3193 & 4436.3770 & -9.2662 & 1.0293 & 0.0094 & 0.5150 & 0.0007\\
6857.5388 & -25625.2710 & 1.7446 & 4421.9564 & -10.6323 & 1.0678 & 0.0125 & 0.5036 & 0.0009\\
6858.5182 & -25622.6933 & 1.4250 & 4432.2060 & -9.9027 & 1.0034 & 0.0099 & 0.4929 & 0.0007\\
6863.5169 & -25622.6876 & 1.3877 & 4434.8193 & -7.8466 & 1.0297 & 0.0103 & 0.5056 & 0.0007\\
6864.5176 & -25624.8996 & 1.1102 & 4424.7214 & -8.4713 & 0.9533 & 0.0077 & 0.4994 & 0.0005\\
6874.4791 & -25620.6138 & 1.6580 & 4440.2229 & -5.6786 & 1.0463 & 0.0121 & 0.5086 & 0.0009\\
7047.8603 & -25624.7545 & 1.2191 & 4434.2142 & -8.8214 & 1.0933 & 0.0090 & 0.5089 & 0.0006\\
7053.8561 & -25621.7439 & 1.4441 & 4446.4398 & -7.8204 & 1.0895 & 0.0103 & 0.5060 & 0.0007\\
7057.8269 & -25622.0109 & 1.4588 & 4452.3988 & -8.1500 & 1.5121 & 0.0123 & 0.5507 & 0.0008 & Rejected \\
7058.8515 & -25617.6495 & 1.8982 & 4442.7452 & -17.1716 & 1.0415 & 0.0136 & 0.5044 & 0.0010\\
7079.8236 & -25619.7617 & 1.2772 & 4436.6205 & -11.9584 & 1.1289 & 0.0094 & 0.5148 & 0.0007\\
7080.8500 & -25621.6358 & 1.4059 & 4427.7199 & -5.2545 & 1.0410 & 0.0100 & 0.5063 & 0.0007\\
7082.8651 & -25623.5924 & 1.1403 & 4434.4908 & -7.9841 & 1.0022 & 0.0078 & 0.5025 & 0.0006\\
7085.7333 & -25621.7323 & 1.2833 & 4430.9624 & -8.7143 & 1.0902 & 0.0098 & 0.5114 & 0.0006\\
7114.8209 & -25625.5802 & 1.4614 & 4410.2902 & -14.3900 & 0.9577 & 0.0102 & 0.5028 & 0.0008\\
7115.7150 & -25627.5907 & 1.3311 & 4428.5220 & -8.7149 & 1.0657 & 0.0092 & 0.5075 & 0.0007\\
7116.7852 & -25627.3237 & 1.2820 & 4424.2680 & -7.4215 & 0.9395 & 0.0083 & 0.5014 & 0.0007\\
7142.7719 & -25625.3593 & 1.2290 & 4428.5104 & -8.7045 & 0.9850 & 0.0083 & 0.5065 & 0.0006\\
7147.7808 & -25616.3909 & 1.3538 & 4441.0481 & -9.6390 & 0.9792 & 0.0103 & 0.5038 & 0.0007\\
7148.7468 & -25620.8954 & 1.2741 & 4429.2924 & -10.0502 & 0.9918 & 0.0094 & 0.5001 & 0.0006\\
7202.5939 & -25616.1481 & 1.6565 & 4441.1922 & 5.6209 & 0.9188 & 0.0097 & 0.4982 & 0.0008\\
7204.6007 & -25623.6987 & 1.4010 & 4452.3792 & 2.2721 & 0.9045 & 0.0081 & 0.4978 & 0.0006\\
7211.5712 & -25624.5994 & 1.3200 & 4447.6138 & 2.6096 & 0.8998 & 0.0075 & 0.4998 & 0.0005\\
7212.6084 & -25624.9280 & 5.6243 & 4437.5059 & -12.1776 & 0.4064 & 0.0213 & 0.5013 & 0.0024\\
7214.5883 & -25625.9395 & 2.0090 & 4454.7852 & 5.6477 & 0.8619 & 0.0135 & 0.5093 & 0.0010\\
7238.5220 & -25620.5551 & 1.4908 & 4444.8883 & 4.9186 & 0.9594 & 0.0104 & 0.5031 & 0.0006\\
7249.4828 & -25623.8234 & 1.8006 & 4447.2116 & 4.2354 & 0.8713 & 0.0112 & 0.4972 & 0.0008\\
7448.8620 & -25628.4475 & 1.4117 & 4457.4186 & 6.7720 & 0.9810 & 0.0103 & 0.5065 & 0.0007\\
7473.8467 & -25621.6222 & 1.1138 & 4455.5312 & 0.5160 & 0.9392 & 0.0082 & 0.5075 & 0.0005\\
7476.8649 & -25621.8943 & 1.1800 & 4459.2948 & 3.4832 & 0.9172 & 0.0083 & 0.5028 & 0.0006\\
7508.4799 & -25624.9467 & 1.0813 & 4467.1596 & -5.9534 & 1.0357 & 0.0107 & 0.5091 & 0.0005 & HARPS-N\\
7508.5698 & -25621.9227 & 1.0769 & 4464.6067 & -7.1361 & 1.2192 & 0.0108 & 0.5287 & 0.0005 & HARPS-N\\
7509.4759 & -25625.1223 & 1.2681 & 4460.0000 & -9.6134 & 0.9486 & 0.0111 & 0.4974 & 0.0005 & HARPS-N\\
7509.5684 & -25626.8176 & 1.4120 & 4462.4128 & -8.3783 & 0.9415 & 0.0128 & 0.4969 & 0.0006 & HARPS-N\\
7510.4709 & -25621.9484 & 1.3119 & 4461.3527 & -5.7705 & 0.9028 & 0.0117 & 0.4927 & 0.0006 & HARPS-N\\
7510.5488 & -25620.3344 & 1.2794 & 4467.5365 & -5.4813 & 1.0562 & 0.0124 & 0.5053 & 0.0006 & HARPS-N\\
7535.4250 & -25625.4364 & 2.1712 & 4464.4853 & -2.7902 & 0.9369 & 0.0239 & 0.5095 & 0.0012 & HARPS-N\\
7536.4320 & -25624.6383 & 1.1442 & 4463.1872 & -8.4287 & 0.9829 & 0.0134 & 0.5065 & 0.0006 & HARPS-N\\
7537.4339 & -25619.0095 & 1.3012 & 4461.0889 & -10.0375 & 0.9686 & 0.0148 & 0.5083 & 0.0007 & HARPS-N\\
7537.5223 & -25618.7979 & 1.1950 & 4464.6137 & -6.3194 & 1.0328 & 0.0149 & 0.5105 & 0.0007 & HARPS-N\\
7538.4145 & -25618.5006 & 1.0871 & 4459.3290 & -6.8755 & 0.9491 & 0.0118 & 0.5029 & 0.0006 & HARPS-N\\
7538.5184 & -25615.3666 & 1.0356 & 4470.6070 & -7.3075 & 0.9111 & 0.0099 & 0.5017 & 0.0005 & HARPS-N\\

 \hline

\end {longtable}
\end{appendix}
\label{lastpage}

\end{document}